\RequirePackage{fix-cm}
\documentclass[twocolumn,epjc3]{svjour3}  
\smartqed  
\RequirePackage{graphicx}
\journalname{Eur. Phys. J. C}
\usepackage{amssymb}
\usepackage{amsmath}

\def\be{\begin{eqnarray} &&}
\def\nonu{\nonumber \\ &&}
\def\ee{\end{eqnarray}}
\def\beq{\begin{equation}}
\def\eeq{\end{equation}}

\usepackage{color}

\def\psla{\rlap \slash}
\def\bew{\begin{widetext}}
\def\ew{\end{widetext}}
\begin{document}

\title{
Pion Generalized Parton Distributions within a fully 
covariant
constituent quark model
}

\author{ Cristiano Fanelli \thanksref{e1,addr1}\and Emanuele Pace
 \thanksref{e2,addr2}\and  Giovanni Romanelli
\thanksref{e3,addr3}\and Giovanni Salm\`e \thanksref{e4,addr4}\and 
 Marco Salmistraro  \thanksref{e5,addr5,addr6}\and
}
\thankstext{e1}{e-mail: cfanelli@mit.edu}
\thankstext{e2}{e-mail: emanuele.pace@roma2.infn.it}
\thankstext{e3}{e-mail: giovanni.romanelli@stfc.ac.uk}
\thankstext{e4}{e-mail: salmeg@roma1.infn.it}
\thankstext{e5}{e-mail: marco.salmistraro21@gmail.com}
\institute{Massachusetts Institute of Technology, Laboratory for Nuclear Science,
 77 Massachusetts Ave, Cambridge,  MA 02139, USA
 \label{addr1}\and 
Phys. Dept. ''Tor Vergata'' University and INFN Sezione
	      di Tor Vergata, Via della Ricerca Scientifica 1, 00133 Rome, 
	      Italy \label{addr2}\and
	      STFC, Rutherford-Appleton Lab., Harwell Campus Didcot OX11 0QX,  UK
	      \label{addr3}\and
	      Istituto  Nazionale di Fisica Nucleare, 
	      Sezione di Roma, P.le A. Moro 2,
 I-00185 Rome, Italy \label{addr4}
 \and    Phys. Dept.   ''La Sapienza'' University,  P.le A. Moro 2,
 I-00185 Rome, Italy      \label{addr5}  \and
 Present addr.:  I.I.S. G. De Sanctis, Via Cassia 931, 00189 Rome, Italy\label{addr6}}

\maketitle

\begin{abstract}
We extend the investigation of the Generalized Parton Distribution 
for a charged pion  within a fully covariant constituent quark
model, in two respects: (i) calculating the tensor distribution  
and (ii) adding the treatment of the 
evolution, needed for achieving a meaningful comparison with both the
experimental parton distribution and the lattice evaluation of the so-called
generalized form factors. Distinct features of our phenomenological covariant quark model 
are: (i) a  4D Ansatz for 
the  pion Bethe-Salpeter amplitude, to be used in  
  the 
Mandelstam formula
for matrix elements of  the relevant current operators, and  (ii) only two 
parameters,  namely a
quark mass assumed to hold  $m_q=~220$ MeV and a free parameter fixed
through the value of the pion decay constant. 
The possibility  of increasing the dynamical content of our covariant constituent
quark model is   briefly discussed in the context of the Nakanishi integral
representation of the Bethe-Salpeter amplitude.

\end{abstract}

\section{Introduction}
\label{intro}
The  present theory of  strong interaction, the Quantum Chromodynamics (QCD),
should in principle  allow one to achieve 
 a complete 3D description of  hadrons, in terms of the Bjorken variable $x_B$
 and the transverse momenta of the constituents. As it is well-known, the needed non
 perturbative description still represents a challenge, that motivates 
  a large amount of valuable efforts, both on the experimental side 
  (gathering new
  accurate data, that in turn impose stringent constraints on theoretical
  investigations)
    and the theoretical one
  (performing  more and more refined  lattice calculations and 
   elaborating more and more  reliable phenomenological models).

Heuristically, while the short-distance behavior of the hadronic state has been
well understood, given the possibility of applying a  perturbative approach, 
entailed by the  asymptotic
freedom,  the long-range part of the hadronic state, that is governed by the
confinement, requests non perturbative tools, suitable for a highly non linear
dynamics.  Coping with the difficult task to gain information on the hadronic
state, in the whole range of its extension, has been the main motivation
 for  elaborating 
phenomenological models, that in general play a helpful role in shedding 
light onto the non
perturbative regime.

Among the phenomenological approaches,  covariant constituent quark models
(CCQMs)  represent an important  step forward, since they exploit a
 quark-hadron
vertex fulfilling the fundamental
property of  covariance with respect to the Poincar\'e group. Moreover, 
CCQM's based on the  Light-front (LF) framework, introduced
by Dirac in 1949 \cite{Dirac}, with
variables defined by: $a^\pm=a^0\pm
a^3$ and  ${\bf a}_\perp\equiv \{a_x,a_y\}$,  appear to be  quite
suitable  for describing  relativistic, interacting systems, like
hadrons. Indeed, the LF framework has  several appealing features 
(see, e.g., \cite{Brodpr}), quite useful 
for  exploring  nowadays issues in hadronic phenomenology.  Beyond the
well-known  fact that the dynamics onto the light-cone is
naturally described in terms of  LF variables, one should
mention : (i) the straightforward separation of the global motion from the
intrinsic one (related to the subgroup property of the LF boosts), (ii) the
largest number of kinematical (i.e. not affected by the interaction) Poincar\'e
generators, (iii)  the large extent of  triviality of the
vacuum, within a LF field theory \cite{Brodpr} (with the caveat of
the zero-mode contributions).
 In particular,
 for the pion, one
can construct the following meaningful Fock expansion onto the null-plane
$$| \pi \rangle =  ~|q\bar{q} \rangle ~~ + 
|q \bar{q} ~ q \bar{q}\rangle ~ + ~
|q \bar{q} ~g\rangle .....$$
where $|q\bar{q} \rangle$ is the valence component. 
It has to be recalled  that an appealing feature of  our approach, based on a covariant description of the quark-pion vertex
(see \cite{Frede09,Frede11} and references quoted therein), is the possibility
of naturally taking into account
contributions beyond the valence term. 

The experimental efforts are very intense for
singling out quantities that are sensitive to the dynamical features of
the hadronic states. In particular, 
in  the last decade, it has been  
 recognized that 
a wealth of information on the 3D partonic structure of hadrons is contained in 
 the Generalized Parton
Distributions (GPDs) (see, e.g., Ref. \cite{Diehlpr} for a general presentation),  as well as  in 
the Transverse-momentum 
Distributions (TMDs) (see,
e.g., Ref.  \cite{Baronpr} for a detailed discussion).
GPDs can be experimentally investigated through the 
Deeply Virtual
Compton Scattering (DVCS), while  TMDs can be studied through   
Semi-inclusive Deep Inelastic 
Scattering (SIDIS) processes, which notably involve polarization degrees of freedom.

Our aim, is to provide a  phenomenological model,  that has the following main ingredients: (i)  a  4D Ansatz for 
the  pion Bethe-Salpeter amplitude, and (ii)   
  the generalization of the 
Mandelstam formula \cite{Mandel}
for matrix elements of  the relevant current operators (notice that  the pion Bethe-Salpeter amplitude is needed in this formula).
Remarkably, we introduce only two 
parameters,  namely a constituent
quark mass   and a free parameter fixed
through the value of the pion decay constant.
 Through  our model, we investigate
the pion state by   thoroughly comparing the results with both experimental and
lattice data relevant for the 3D description of the pion. In this
paper, we  complete the evaluation of  the leading-twist pion GPDs,  
calculating  the so-called tensor GPD (see Ref. \cite{Frede09} for
 the
vector GPD and Ref. \cite{Pace12,Pace13} for preliminary calculations of the tensor one). Moreover, 
in order to accomplish the previously mentioned comparisons, we consider  
 the evolution of quantities that can be
extracted from the GPDs, like the parton
distribution function (PDF) and the generalized form factors (GFF). We anticipate that
only the   leading order (LO) evolution has been implemented by using  
the standard code of Ref. \cite{Kumano}.
 In particular, the comparison has been performed between our LO results and 
  the experimental pion PDF extracted in Ref. 
\cite{Conway} 
(see Ref. \cite{Holt} for the NLO extraction)  and  the available 
lattice calculations of GFFs as given in Refs.
\cite{Bromth,Lat07,Brom08,Brom08b}.

The paper is organized as follows. In Sec. \ref{general}, the general formalism and the definitions
are briefly recalled. In Sec. \ref{CCQM}, our Covariant Constituent Quark Model is presented. In Sec. \ref{evol}, 
the LO evolution of the quantities we want to compare is
thoroughly discussed, with a particular care to the determination of the initial scale of our model.
In Sec. \ref{results}, the comparison of our results with  both the experimental PDF and the available lattice
calculations is presented. Finally in Sec. \ref{concl}, the Conclusion are drawn.

\section{Generalities}
\label{general}
In this Section, the physical quantities, GPDs and TMDs,  that allow us to achieve 
a detailed 3D description of
a pion are  shortly introduced, since they represent  the target of
the investigation within our CCQM (for  accurate and extensive reviews
on GPDs, see, e.g. \cite{Diehlpr} and on TMDs, see, e.g. \cite{Baronpr}). 
 For the pion, given its null
total angular momentum, one has two GPDs  and
 two TMDs, at the leading twist.
\subsection{Generalized Parton
Distributions}
\label{formalism}
As it is well known, GPDs 
are    LF-boost invariant functions, and
allow one to parametrize matrix elements (between hadronic states) 
 involving quark and gluon fields.
In particular, 
GPDs are off-diagonal (respect to the hadron four-momenta, i.e.
 $p_f\ne p_i$,) matrix elements of  quark-quark (or gluon-gluon) 
correlator projected onto the  Dirac basis (see,
e.g., Ref. \cite{Meis08} for a thorough investigation of the pion case).
The appealing feature of GPDs is given by the ability of summarizing
in a natural way   information contained  in several
observables investigated in different kinematical regimes, like   
electromagnetic (em), form factors (FFs) or  
PDFs. 

The pion 
 has two  leading-twist quark GPDs: i) the vector, or no spin-flip, GPD,
 $H^I_\pi(x,\xi,t)$,
 and ii) the tensor, or
spin-flip, GPD, $ E^I_{\pi,T}(x,\xi,t)$ (where $I=IS,IV$ labels 
 isoscalar
and isovector GPDs, respectively).
In order to avoid
 Wilson-line contributions, one
  can choose the light-cone gauge \cite{Diehlpr}
 and get 
\be
2~\left(\begin{array}{c} 
 H^{IS}_{\pi}(x, \xi, t) \\ ~\\ H^{IV}_{\pi}(x, \xi, t) \end{array}\right)
 =
  \int {dk^- d{\bf k}_\perp\over 2}\nonu
  \times
 \int \frac{dz^- dz^+ d{\bf z}_\perp}{2(2 \pi)^4} 
 e^{i[(xP^+z^-+k^-z^+)/2 -{\bf z}_\perp \cdot {\bf k}_\perp]}
 \nonu \times ~\langle p'| 
\bar{\psi}_q (-\frac12z) \gamma^+ \left(\begin{array}{c} 1 \\ ~\\\tau_3
  \end{array}\right) \psi_q(\frac12z) 
|p \rangle
 = 
 \int \frac{dz^- }{4 \pi}\nonu \times ~ e^{i(xP^+z^-)/2} ~\langle p'| 
\bar{\psi}_q (-\frac12z) \gamma^+ \left(\begin{array}{c} 1 \\ ~\\\tau_3
  \end{array}\right) \psi_q(\frac12z) 
|p \rangle\big|_{\tilde z=0} \nonu
\label{vector}\ee
and
\be
\frac{P^+ \Delta^j- P^j\Delta^+}{P^+ m_\pi}\left(\begin{array}{c} 
 E^{IS}_{\pi T}(x, \xi, t) \\ ~\\E^{IV}_{\pi T}(x, \xi, t) \end{array}\right)
 =\int \frac{dz^- }{4 \pi}
 \nonu \times ~ e^{i(xP^+z^-)/2}~\langle p'| 
\bar{\psi}_q (-\frac12z) ~i\sigma^{+j} ~\left(\begin{array}{c} 1 \\ ~\\\tau_3
  \end{array}\right) \psi_q(\frac12z) 
|p \rangle\big|_{\tilde z=0} \nonu
\label{tensor}\ee
where $\tilde z \equiv \{z^+ = z^0 + z^3 , {\bf z}_\perp\}$,
  $\psi_q(z)$ is the  quark-field  isodoublet and the standard GPD variables are
  given by
  \be
  x={k^+\over P^+}~~, \quad \xi= ~-~{\Delta^+\over 2 P^+}~~, \quad 
  t=\Delta^2 ~~, \nonu  \Delta=p'-p~~, \quad  P= {p'+p\over 2}
  \ee
  with the initial
   LF momentum of the active quark equal to
   $\{k^+- \Delta^+/2,{\bf k}_\perp-{\bf \Delta}_\perp/2\}$. The factor 
   of two multiplying
   the vector GPD is chosen for normalization purpose, so that for a charged
   pion one has
   \be
   F_\pi(t)= \int_{-1}^1 dx ~ H^{IV}_\pi (x, \xi, t)= \int_{-1}^1 dx ~
  H^{u}_\pi (x, \xi, t)
 \label {emff}\ee
 where $H^u_\pi=H^{IS}_\pi + H^{IV}_\pi$, and $H^{IS}_\pi$ is odd in $x$ while $H^{IV}_\pi$ is even (see, e.g.
 \cite{Frede09}). Finally, it is useful for what follows
 to recall the relation with 
  the parton distributions,
  $q(x)$,  viz
  \be
  H^u_\pi (x, 0,
 0)=\theta(x)u(x)-\theta(-x) \bar u(-x)~~.\ee
 
 At the present stage, only a few moments of the pion GPDs have been evaluated
 within lattice QCD, but they represent a valuable test ground for any
 phenomenological model that  aspires to yield meaningful insights into the pion
 dynamics. In view of the numerical results discussed below,
   we briefly recall how the Mellin moments can be covariantly  parametrized
    through  the GFFs, that
    are the quantities  adopted for comparing
      lattice calculations and
    phenomenological results.

  The relation between the non-spin flip GPD and the em FF given in Eq.
  \eqref{emff}
 for a charged pion
 can be in some sense generalized, if one considers  Mellin moments of both vector and tensor
 GPDs. Then one  obtains the corresponding  GFFs. For instance, one can write 
 the following 
 Mellin moments of both  vector and tensor GPDs for the 
  $u$-quark (see \cite{Diehlpr,Hagelpr} for a review)
\be
 \int_{-1}^1 dx \, x^n H^u_\pi (x, \xi, t) = 
\sum _{i=0}^{[(n+1)/2]} (2 \xi) ^{2i} A^u_{n+1,2i} (t)~~, 
\label{VecGFF}\\ &&
\int_{-1}^1 dx \, x^n E^u_{\pi,T} (x, \xi, t) = 
\sum _{i=0}^{[(n+1)/2]} (2 \xi) ^{2i} B^u_{n+1,2i} (t)  
\label{TGFF}\ee 
where  the symbol $[...]$ indicates the integer part of the argument.   
 In Eqs. \eqref{VecGFF} and  \eqref{TGFF},  $A^u_{n+1,2i}(t)$ is  a  vector GFF for a $u$-quark and   
 $B^u_{n+1,2i}(t)$ a tensor GFF, respectively. It is worth noting that one can introduce a
 different decomposition in terms of isoscalar and isovector components instead
 of a flavor decomposition. In particular,  if $n+1$ is  even (odd)
  one has an isoscalar (isovector) GFF. A striking feature is shown by the rhs of Eqs. \eqref{VecGFF} and 
  \eqref{TGFF}, the so-called polinomiality, i..e. the dependence upon finite
  powers of the variable $\xi$.  This polynomiality property follows 
  from completely general properties like covariance, parity 
  and time-reversal invariance; for this reason it can be a good test for 
  any model.

   By considering the first vector and tensor moments one gets the following important
   relations 
 \be \int_{-1}^1 dx \,  H^u_\pi (x, \xi, t)= A^u_{1,0} (t)=F_\pi(t)
 \label{ff}\ee
 and
 \be\int_{-1}^1 dx \,  E^u_{\pi,T} (x, \xi, t)= B^u_{1,0} (t)\ee
 where $B^u_{1,0} (0)\ne 0$ is  the tensor charge for $n=0$, also called
   tensor anomalous magnetic moment (see Ref. \cite{Hagelpr}).
 Notably,  Eq. \eqref{VecGFF} leads to the following relation involving the
  Mellin moments of the 
 PDF and  $A^u_{n+1,0} (0)$, viz
 \be
 <x^n>_u=\int_{-1}^1 dx \, x^n H^u_\pi (x, 0, 0) =  A^u_{n+1,0} (0)
 \label{normA}
 \ee

A physical interpretation of GFFs (see, e.g., \cite{Bur,Bur05,Bur08})  can be achieved by properly
generalizing  the standard interpretation of the non relativistic em FFs 
to a
relativistic framework. Non relativistically,  the em FFs are the  3D Fourier transforms 
of
 intrinsic (Galilean-invariant)  
em distributions  in the coordinate space (e.g., for the pion, 
one has the charge distribution, while, for the nucleon, one has both charge and
magnetic distributions). In the relativistic case, one should   consider  Fourier 
transforms of GPDs, that depend
   upon variables 
  invariant under LF boosts. Indeed, only the transverse part of $\Delta^\mu$
can be trivially conjugated to  variables in the coordinate space, while for  $x$ 
and $\xi$ (proportional to
$\Delta^+$) this is not possible. Therefore,  keeping  the description invariant for  
proper boosts (i.e. LF boosts), one can introduce 2D
 Fourier transforms with
respect to ${\bf \Delta}_\perp$.  Such a Fourier transform  allows one to  investigate  the spatial
distributions of the quarks in the so-called impact-parameter space (IPS).
  In particular, from Eq. \eqref{VecGFF} and  \eqref{TGFF}, it straightforwardly follows
that, for $\xi=0$, only $A^u_{n+1,0} (\Delta^2)$ 
  and 
$B^u_{n+1,0} (\Delta^2)$ survive. Due to the LF-invariance of $\xi$, one has 
 an infinite set of frames (Drell-Yan frames) where 
$\xi=0$. In these frames, where 
  $\Delta^+=0$ and  
$\Delta^2=-\Delta^2_\perp$,  one can introduce 
the above mentioned 2D Fourier transforms in a boost-invariant way (recall that,
for a given
reaction,  the final state or both final and initial states have to be boosted). 
One can write
\be
\tilde A^q_{n}(b_\perp)= \int {d {\bf \Delta}_\perp\over (2 \pi)^2} ~e^{i 
{\bf\Delta}_\perp \cdot {\bf b}_\perp}
  A^q_{n,0}(\Delta^2)~, \nonu
\tilde B^q_{n}(b_\perp)= \int {d {\bf \Delta}_\perp\over (2 \pi)^2}
~ e^{i {\bf  \Delta}_\perp \cdot {\bf b}_\perp} 
B^q_{n,0}(\Delta^2) 
\ee
where $b_\perp=|{\bf b}_\perp |$, is the impact parameter.
In general, the Fourier transform of GFFs, for $\xi=0$,  yield
 quark densities 
in the IPS \cite{Bur,Bur05,Bur08}. 
In particular,  $\tilde A_{n}(b_\perp)$ represents the probability density  
of finding an unpolarized quark 
in the pion at a certain distance $b_\perp$ from the transverse
center of momentum. In addition, if one  considers the polarization degrees of
freedom, then  one introduces 
  the probability density of finding 
 a quark with a given transverse polarization, 
 ${\bf s}_\perp$  in a certain Drell-Yan frame. In the IPS, such a probability distribution
 is 
\be
\rho^q_n ({\bf b}_\perp,{\bf s}_\perp) = 
\frac12 \left [ \tilde A^q_n(b_\perp) + 
\frac {s^i\epsilon ^{ij} b^j}{b_\perp} ~\Gamma^q_n(b_\perp)
\right ] 
\label{rhodef}\ee
where
\be
\Gamma^q_n(b_\perp) = -\frac{1}{2 m_{\pi}}\frac{\partial ~\tilde B^q_n(b_\perp)}{\partial ~b_\perp}
\label{gamdef}
\ee
 It is worth noting that  the quark longitudinal (or helicity) distribution density 
 is given only by    the first term in Eq.
\eqref{rhodef},
since the pion is a pseudoscalar meson and   the term $\gamma_5 \psla
s_L$  in the quark density operator has a vanishing expectation value, due to
 the parity invariance \cite{Brom08b,Diehl05}.

Equation \eqref{rhodef} is quite rich of information and clearly indicates the
pivotal role of GPDs for accessing 
  the quark distribution
in the IPS. Moreover, as a
closing remark, one could   
exploit  the spin-flip GPD $E^q_{\pi T}$  to extract  more elusive information on the quasi-particle nature
of the constituent quarks, like their possible anomalous magnetic moments,
once the vector current that governs the quark-photon coupling  is suitably 
improved (see subsect. \ref{gffevol} and Ref. \cite{Brom07} for a discussion within the lattice 
framework).
  \subsection{Transverse momentum distributions}
   TMDs are  diagonal (in the pion
   four-momentum\footnote{Notice that $\Delta^\mu
=(p^\pi_f-p_i^\pi)^\mu=0$ leads to 
 $\xi=t=0$. }) matrix elements of  the  
   quark-quark (or gluon-gluon) 
correlator  with the proper Wilson-line contributions (see,
e.g., Ref. \cite{Meis08}) and  suitable Dirac structures.  Moreover,
 TMDs depend upon  $x$
and the  quark transverse momentum, ${\bf k}_\perp$, that is not the conjugate of ${\bf b}_\perp$. It should be pointed out that  in general the
Wilson-line effects must be carefully analyzed, due to the explicit dependence
 upon 
 ${\bf k}_\perp$  (recall that for GPDs such dependence is integrated out).
   At the leading-twist, one
has two TMDs, for the pion: the T-even $f^q_{1}(x, |{\bf k}_\perp|^2)$,
that yields the probability distribution to find an unpolarized quark with LF momentum
$\{x,{\bf k}_\perp\}$ in the pion, and the  T-odd  
$h^{q\perp}_1(x, |{\bf k}_\perp|^2,\eta)$, 
related to a  transversely-polarized quark and called {\em Boer-Mulders}
distribution \cite{Boer}. 

The two TMDs allow one to parametrize the distribution
of a quark with given  LF momentum and transverse polarization, i.e.
(see, e.g., Ref. \cite{Brom08,Meis08})
\be
\rho^q (x,{\bf k}_\perp,{\bf s}_\perp,\eta) = \nonu =
\frac12 \Bigl [f^q_{1}(x, |{\bf k}_\perp|^2)  + 
{s^i\epsilon ^{ij} k_\perp^j \over m_\pi}
 ~h^{q\perp}_1(x, |{\bf k}_\perp|^2,\eta)
\Bigr ] 
\label{rhodefk}
\ee
where the dependence upon the variable $\eta$ in $h^\perp_1$ is generated by  
the    Wilson-line effects, whose role is essential for investigating a non
vanishing $h^\perp_1$ (see e.g. \cite{Boer}).

At the lowest order,  the unpolarized TMD 
$f^{q}_{1}$,  is given by the proper combination of the
isoscalar and isovector components, that are defined by
\be
2\left(\begin{array}{c} 
 f^{IS}_{1}(x, |{\bf k}_\perp|^2) \\ ~\\  f^{IV}_{1}(x, |{\bf k}_\perp|^2) \end{array}\right)
 = 
 \int {dz^-  d{\bf z}_\perp \over 2 (2\pi)^3}e^{i [xP^+z^-/2 -
 {\bf k}_\perp\cdot
 {\bf z}_\perp] }
 \nonu \times ~\langle p| 
\bar{\psi}_q (-\frac12z) \gamma^+ \left(\begin{array}{c} 1 \\ ~\\\tau_3
  \end{array}\right) \psi_q(\frac12z) 
|p \rangle\big|_{ z^+=0}~~~,
\ee
After integrating over ${\bf k}_\perp$, one gets  
  the standard unpolarized parton
distribution $q(x)$, viz
\be
q(x)= \int d{\bf k}_\perp~ f^{q}_{1}(x, |{\bf k}_\perp|^2)= ~
H^{q}_{1}(x, 0,0)~~~.
\label{untmd}\ee
The T-odd TMD, $h^\perp_1(x, |{\bf k}_\perp|^2,\eta)$ needs a more careful
analysis, since it 
  vanishes at the lowest order in perturbation
theory.  As a matter of fact,  it becomes proportional to the matrix elements
\be
\langle p| 
\bar{\psi}_q (-\frac12z) ~i~\sigma^{+j}~ \left(\begin{array}{c} 1 \\ ~\\\tau_3
  \end{array}\right) \psi_q(\frac12z) 
|p \rangle\big|_{ z^+=0}~~~,
\ee
 that are equal to zero, due to  the time-reversal invariance.
In order to get a non vanishing Boer-Mulders distribution,
one has to  evaluate at least a first-order correction,  involving Wilson 
lines 
  (see, e.g., Refs. \cite{Lu04} and  \cite{Meis08}). Moreover, by adopting the light-cone gauge and the advanced boundary condition 
for the gauge field, the effect of the Wilson lines (final state interaction
effects) can be  shifted into complex phases affecting the initial state 
(see, e.g., Ref. \cite{Beli03}).

\section{The Covariant Constituent Quark Model}
\label{CCQM}
The main ingredients of our covariant constituent quark model  are 
 two: i)  the extension to the  GPDs and TMDs of 
  the  Mandelstam formalism \cite{Mandel}, originally introduced for
  calculating   matrix elements of the em current
  operator when a relativistic  interacting system is investigated, and ii) a  model of
 the 4D quark-hadron  vertex, or equivalently  the Bethe-Salpeter
amplitude,  necessary for applying the Mandelstam approach. 
In particular, we have assumed a pion  Bethe-Salpeter amplitude (BSA) 
with the following form
\be
\Psi (t,p) =  -{m \over f_\pi}~S\left(t+p/2\right)~  \Gamma(t,p) 
~
S\left(t-p/2\right) 
\nonu\label{bsa} 
\ee
where $p=p_q+p_{\bar q}$ is the  total 
momentum, $t=(p_q-p_{\bar q})/2$  the relative momentum of the $q\bar q$ 
pair (by using the four-momenta  $k$, $\Delta$ and $P$ previously introduced, 
one has  $t+p/2=k-\Delta/2$, 
 and
$t-p/2=k-P$). In Eq. \eqref{bsa}, $S(p_q)=1/(\psla p_q-m_q+\imath \epsilon)  $ is 
 the fermion propagator and  $\Gamma(t,p)$  the quark-pion vertex. In the present work,
only the dominant Dirac structure has been assumed, viz
\be
\Gamma(t,p)=\gamma_5~\Lambda_\pi(t,p)
\label{gampi}\ee
with $\Lambda(t,p)$ a suitable momentum-dependent scalar function that contains the dynamical information
(see the following subsections
for more details). Indeed, 
 Dirac structures contributing to $\Gamma(t,p)$ beyond $\gamma^5$ should be taken into
 account, but  they have a minor  impact  on the pion BSA, as thoroughly discussed in   Ref. \cite{maris}. 

For the sake of completeness, let us recall that the quark-pion vertex  fulfills 
  the homogeneous BS equation  that reads as follows
 \be
 \Gamma(t,p) =\nonu= \int {d^4t' \over (2 \pi)^4}
~{\cal K}(t,t')~S\left(t'+p/2\right)~  \Gamma(t',p) 
~
S\left(t'-p/2\right)
 \label{bse}
\nonu \ee
where ${\cal K}(t,t')$ is the kernel given by the infinite sum of irreducible diagrams (see, e.g.,
\cite{IZ}). 

Finally, it is important to emphasize that our investigation,  based on a covariant
 description of the quark-pion vertex, naturally goes beyond 
 a purely valence description of the pion \cite{Frede09,Frede11}.

\subsection{The Mandelstam Formula for the electromagnetic current} 
The Mandelstam formula allows one to express  the
matrix elements of the em current of a composite bound system, within a field
theoretical approach \cite{Mandel}.  It has been applied for evaluating
the FFs of both  pion \cite{Melo02,Melo04,Melo06}
and nucleon \cite{Melo09}, obtaining a nice description of both space- and
timelike FFs. Furthermore, it has been  exploited for calculating the
vector GPD of the pion  \cite{Frede09,Frede11} and for a preliminary 
evaluation of the
tensor GPD \cite{Pace12,Pace13}.

 For instance, in the case of the em spacelike FF of  the pion,
  the Mandelstam formula,  where  the quark-pion vertex given in Eq.
 \eqref{gampi} is adopted, reads
  (see, e.g., Ref. \cite{Melo02,Melo04,Melo06})
\be
j^{\mu} ~=
 ~-\imath  e ~{\cal R}\nonu \times~
 \int
\frac{d^4k}{(2\pi)^4}
 \Lambda_{\pi}(k+\Delta/2,p') 
\Lambda_\pi(k-\Delta/2,p) \nonu \times~  
Tr[S(k-P)\gamma^5
S(k+\Delta/2)V^\mu(k,q)S(k-\Delta/2)  \gamma^5  ]  
\nonu\label{current1}\ee
where ${\cal R}=2 N_c m^2_q/f^2_\pi$, $f_\pi$  is the pion decay constant
 $N_c=3$  the  number of
colors, $m_q$  the CQ mass   and 
$V^\mu(k,q)$  the quark-photon vertex, that  we have simplified to
$\gamma^\mu$ in the spacelike region. In presence of a CQ, one
could  add to the bare  vector current  a term proportional to an anomalous magnetic moment, namely a term
like $$ i{\kappa_q \over 2 m_q}\sigma^{\mu\nu} \Delta_\nu~~~,$$
as in Ref. 
\cite{Brom07} (where it has been adopted 
 an improved vector current  within a lattice framework).
 It should be pointed out that  the above mentioned anomalous magnetic moment is not used in
the present work.
 
Within CCQM, the expression of the decay constant in term of $\Lambda_\pi$ 
reads (cf Ref. \cite{Melo02})
 \be
f_\pi=- i\frac{m_q}{f_\pi}{N_c\over m^2_\pi}
\int\frac{d^4k}{(2\pi)^4}~\Lambda_\pi(k-\Delta/2,p) \nonu \times ~
Tr \Bigl[ \psla p \gamma_5 S(k-\Delta/ 2) \gamma_5 S(k-P)
\Bigr]=
\nonu =i \frac{m_q^2 ~N_c}{(2\pi)^2  f_\pi}
\int {d\kappa^+ \over \kappa^+~(\kappa^+-m_\pi)}
~\int d^2 \kappa_{\perp}    
\Phi(\kappa^+, \kappa _\perp)
 \label{fpi}
 \ee
 where $\kappa=k-\Delta/2=k-P +p$ (recall $p^2=m^2_\pi$), $\kappa_\perp=|{\bf \kappa}_\perp|$ and 
 \be
 \Phi(\kappa^+, \kappa _\perp) = \kappa^+~(\kappa^+-m_\pi)
 \int {d\kappa^- \over 2\pi} \nonu \times{\Lambda_\pi(\kappa,p) \over
  \Bigl[\kappa^2 - m^2_q 
  +i\epsilon\Bigr]~\Bigl[(\kappa-p)^2 - m^2_q 
  +i\epsilon\Bigr]}~
  \label{DA1}
\ee
is the valence wave
 function.
It should be recalled that $\Phi$, properly integrated on $ \kappa _\perp$,
yields the pion distribution amplitude (DA) (see Eq. \eqref{DA2} and, e.g.,
 Ref. \cite{Diehlpr} for a general
discussion on the DAs and their evolution).
The generalization of Eq. \eqref{current1} to the case of GPDs, can be found in 
Ref. \cite{Frede09,Frede11}
for the vector GPD, and in \cite{Pace12,Pace13} for the tensor one, but 
for the sake of completeness, let us give the expression of
 both vector and tensor GPDs   for the $u$ quark, viz
\be
 2 ~H^u(x,\xi,t) = -\imath {\cal R} ~
\int
\frac{d^4k}{(2\pi)^4} ~ \delta[P^+x-k^+] \nonu \times~
 \Lambda(k-P,p^{\prime})\; \;
\Lambda(k-P,p)\nonu \times ~  
Tr\left [ S\left({k}-{P}\right)
\gamma^5 S\left({k}+\Delta/2
\right) 
\gamma^+S\left({k}-\Delta/2
\right)\gamma^5\right ]\nonu
 \label{ccqmv}
\ee
and 
\be
{P^+ \Delta^j- P^j\Delta^+ \over  ~ P^+m_\pi}
 E^u_{\pi T}(x, \xi, t) = 
 \nonu =i {\cal R}
\int \frac{d^4 k }{(2 \pi)^4} \delta[P^+x-k^+]~\Lambda(k-P,p')~ \Lambda(k-P,p) \nonu
\times 
 Tr[S({k}-{P})\gamma^5S({k}+\Delta/2)\gamma^+\gamma^j 
 S({k}-\Delta/2)\gamma^5 ]~
\label{ccqmt}\ee
where
$j=1,2$.
The $\delta$ function allows one to have the correct support for the
{\em active} quark, i.e.  when
$|\xi| \le x \le 1$.  This kinematical region corresponds to the so-called
 Dokshitzer-Gribov-Lipatov-Altarelli-Parisi (DGLAP) region 
\cite{dglap1,dglap2,dglap3}, or valence region.
 Moreover,   CCQM  is able  to address also kinematical region beyond 
 the valence one, i.e. $-\xi\le x\le \xi$, given the covariance property. This
 region is called 
 the Efremov-Radyushkin-Brodsky-Lepage 
(ERBL)  region \cite{erbl1,erbl2}, or non valence region. If one  adopts
 a Breit frame with 
$\Delta^+ = -\Delta^- \ge 0$, then the ERBL region can be investigated. As a matter of fact,  in such a frame one can
   access the whole
range of the variable $\xi$, i.e. $-1\leq \xi \leq 1$, and   analyze
both 
valence and non valence regions within the same approach. This allows one to shed
light on 
 the interesting
topic of the smooth transition from the DGLAP (valence) regime to the ERBL (non
valence) one. 

The expression of the unpolarized TMD $ f^q_1(x,|{\bf k}_\perp|^2)$ 
\cite{Frede09}  can be easily obtained from
the integrand of the vector GPD, Eq. \eqref{ccqmv}, by recalling the relation in Eq.
\eqref{untmd}.

 \subsection{The four-momentum dependence of the Bethe-Salpeter amplitude}
 \label{SecBSA}
 As above mentioned, in our CCQM we focus on the main contribution to the pion BSA, 
 i.e. the term
 containing   the Dirac matrix $\gamma_5$. This implies that  we have 
 to consider only one scalar
 function for  describing the dependence upon the four-momenta present 
 in the
 problem. Unfortunately, solutions  of the homogeneous BSE for hadrons  are  still lacking in Minkowski space
 given the extraordinary complexity of QCD, nonetheless 
 very relevant investigations have been carried out  in Euclidean space, within the lattice  framework
 \cite{Bromth,Lat07,Brom08,Brom08b} or combining  BSE and
  Dyson-Schwinger equation (DSE)
 (see, e.g., \cite{Holt10} and references quoted therein)
  or by exploiting a  3D reduction of the BSE itself  
(see, e.g., \cite{Gross14}). On the other hand, since we would carry on 
a comparison with a wide set of data, from both
experiments and  lattice, we resort to adopt  a phenomenological Ansatz, 
that depends remarkably upon   
only two parameters. This  allows us to explore the
potentiality of the Mandelstam approach in capturing the 
main features of the physical quantities under
consideration, while   having a reasonable predictive power, given the small set of
free parameters. 

The following  analytic 
covariant Ansatz for the momentum dependence 
 of the BSA  has been adopted
 \be
 \Lambda(t,p)= \nonu =C~ 
{1 \over \left[(t+p/2)^2-m^2_{R} + \imath \epsilon\right]}
 ~
{1 \over\left[(t-p/2) ^2-m^2_{R}+ \imath \epsilon\right]} 
\nonu
\label{vertexp}
\ee
where 
the parameter $m_R$  is adjusted  to fit   $f_{\pi}$, while 
 the constants  $C$ is fixed through the charge normalization, $F_{\pi}(t=0)=1$,
 that amounts to
 the standard normalization of the BSA, but in impulse approximation.

It is worth noting that the expression in Eq. \eqref{vertexp} can be cast 
(see below) in a form
suggested by the integral representation of the 4D $n$-leg transition 
amplitudes (we are actually interested to the $3$-leg amplitude, i.e. the vertex
$\pi \to
q\bar q$) 
elaborated by 
Nakanishi in the 60's \cite{Naka}, within a perturbation-theory framework. 
To quickly illustrate the appealing features of 
this integral representation, one should consider the $n$-leg transition amplitude
for a many-scalar  interacting system, and the infinite set of Feynman diagrams
contributing to determine the amplitude itself. In this case, it turns out that 
the amplitude 
is given   by the 
folding of a
weight function (called the Nakanishi weight function) and a denominator (with 
some exponent) that
contains all the independent scalar products obtained from the $n$
external four-momenta. It has to be pointed  out that  the analytic behavior of the
amplitude is fully determined by such a denominator, and this clearly makes    the Nakanishi integral
representation a valuable tool for investigating 4D transition amplitudes. For $n=3$, one can apply the
integral representation to   the
vertex function for a system composed by two constituents, and explicitly discuss the 
analytic structure, i.e. the core of the physical content.  Another pivotal motivation
that  increases the interest on  the Nakanishi framework is given by the following computational finding: 
 even if the Nakanishi 
integral representation
has been formally established  by considering the whole infinite
set of the Feynman diagrams contributing to an amplitude, i.e. 
 a perturbative regime, it
has been numerically shown that also in  a non perturbative
framework, like the homogeneous BSE (relevant for describing bound systems), the Nakanishi
representation plays an essential role for obtaining {\em actual} solutions for the
vertex function or, equivalently, for the BSA. Applying the Nakanishi representation as an Ansatz for
 the
solution of the BSE one can  determine the unknown Nakanishi weight function and achieve a genuine 
numerical
solution of the BSE in Minkowski space. This approach has been applied to 
  the 
ladder BSE
 for two-scalar  and two-fermion systems (see, e.g.,Refs.
    \cite{CK06,CK10,Frede12,Frede14,Frede15} for the Nakanishi approach in 
    Minkowski space and, for the sake of 
    comparison, Ref.
     \cite{Dork1,Dork2} for  two-fermion
    systems  within the Euclidean hyperspherical approach), opening a viable path for phenomenological studies within a non
perturbative regime.
  
Within the Nakanishi approach, the vertex function (or three-leg amplitude) can be
 written as follows
\be
\Lambda(t,p)= ~\int_0^\infty d\gamma \int^1_{-1}dz ~{g(\gamma,z;\kappa^2)\over
\Bigl[ \gamma +\kappa^2 -t^2 -z~p\cdot t -i\epsilon\Bigr]^2}\nonu
\label{naka1}
\ee
 where  $\kappa^2=m^2_q -p^2/4$ and $g(\gamma,z;\kappa^2)$ is called Nakanishi weight function.
 If we take $g(\gamma,z;\kappa^2)=\delta(\gamma-m^2_R +m^2_q)$, one obtains 
 Eq. \eqref{vertexp}. 
   It should be pointed out that while waiting for numerical solutions of the two-fermion system with
   more refined phenomenological kernels (for the ladder approximation see Ref.
   \cite{CK10}),  
   one could perform an   intermediate step, still in the realm of 
  Ansatzes, substituting Eq. \eqref{vertexp} with 
   Eq. \eqref{naka1}, but adopting 
   a different choice of the Nakanishi weight function, e.g.
  by substituting the simple delta-like form with  more realistic 
   functions (see e.g.  \cite{Frede14} for the Nakanishi weight functions of a
   two-scalar system obtained by actually solving 
  the homogeneous BSE in the ladder approximation). In order to set the reference line for
   the next steps in the elaboration of our CCQM (presented elsewhere),  we will adopt the very 
   manageable 
  form given in Eq. \eqref{vertexp}, in the following
  comparisons with the experimental and lattice results (see below, Sec. \ref{results}).

 \section{Evolution of  Mellin Moments   and  GFF's}
 \label{evol}
In order to compare our results for PDF and GFFs with experimental data and
lattice calculations, it is fundamental to suitably evolve the CCQM outcomes,
  from the unknown
scale $\mu_{CCQ}$ to the needed ones, namely $\mu_{exp}$ and $\mu_{LAT}$.

Our strategy for determining  an acceptable $\mu_{CCQ}$ is  to study  the evolution 
of the non singlet PDF Mellin 
within a 
LO framework,  considering flavor numbers 
up to  $N_f=4$.
 It should be pointed out that the choice to adopt the  LO framework, seems to be well  
 motivated by  the phenomenological nature
 of the CCQM, and by the present uncertainties still affecting both 
 experimental and lattice GFFs.

Let us shortly summarize  our procedure for assigning a scale $\mu$  to our
calculations. The main  ingredient   to be  considered are the  
 Mellin moments of  the non singlet distribution $f_{NS}(x,\mu)$, viz
\be
M_{NS}(n,\mu)=\int_0^1dx~x^{n}~f_{NS}(x,\mu)
\ee 
where $f_{NS}$ is related  
to the unpolarized GPD, as follows
\be
f_{NS}(x,\mu_{CCQ})=2H^{I=1}(x,0,0)
\label{fns}\ee
Mellin moments evolve from a scale $\mu_0$ to the scale $\mu$ through  
very  
 simple expressions
(see, e.g., \cite{greiner}), that 
 for the 
non singlet, singlet and gluon  moments read
\be
\frac{dM_{NS}(n,\mu)}{dln\mu^2}=\frac{\alpha_s^{LO}(\mu,N_f)}{2\pi}
\frac{\gamma^{(0)}_{qq}(n)}{2\beta_0}~M_{NS}(n,\mu)
\label{MOMNS}\\ &&
\frac{d\overrightarrow{M}(n,\mu)}{dln\mu^2}=\frac{\alpha_s^{LO}(\mu,N_f)}
{2\pi}\frac{\Gamma^{(0)}(n)}{2\beta_0}
~\overrightarrow{M}(n,\mu)\label{MOMS}
\ee
where 
\be
\overrightarrow{M}(n,\mu)=\left(\begin{array}{c}M_S(n,\mu)\\
M_G(n,\mu)\end{array}\right)
\ee
 In Eqs. \eqref{MOMNS} and \eqref{MOMS}, the   LO anomalous dimensions are  indicated
 by $\gamma^{(0)}_{ab}$
 while the $2\times 2$ matrix $\Gamma^{(0)}(n)$  is given by
\be\label{anodimatrix}
\Gamma^{(0)}(n)=\left(\begin{array}{ccc}\gamma_{qq}^{(0)}(n)&~&\gamma_{qG}^{(0)}(n)\\
&\\
\gamma_{Gq}^{(0)}(n)&~&\gamma_{GG}^{(0)}(n)\end{array}\right).
\ee
Let us recall that each anomalous dimension $\gamma^{(0)}_{ab}(n)$ is  
obtained from the corresponding LO splitting function. In particular, for 
the unpolarized case, one has (see, e.g., \cite{greiner})
\be
\gamma_{qq}^{(0)}(n)=-\frac{8}{3}\left[3+\frac{2}{(n+1)(n+2)}-4\sum_{k=1}^{n+1}\frac{1}{k}\right]
\label{anodim1}\\ &&
\gamma_{qG}^{(0)}(n)=-2\left[{n^2+3n+4\over(n+1)(n+2)(n+3)}\right]\label{anodim2}\\ &&
\gamma_{Gq}^{(0)}(n)=-\frac{16}{3}\left[{n^2+3n+4\over n(n+1)(n+2)}\right]\label{anodim3}\\ &&
\gamma_{GG}^{(0)}(n,N_f)=-6\Bigl[{\beta_0(N_f)\over 3}+8~{n^2+3n+ 3\over n(n+1)(n+2)(n+3)}
\nonu -4\sum_{k=1}^{n+1}\frac{1}{k}\Bigr]\label{anodim4}.
\ee
where 
\be\label{beta0}
\beta_0(N_f)=11-\frac{2}{3}N_f
\ee
By taking into account the eigenvalues of $\Gamma^{(0)}(n)$,  given by (see \cite{anodimrad} for details)
\be\label{eigen}
\gamma_{\pm}(n)=\frac{1}{2}\left[\gamma_{qq}^{(0)}(n)+\gamma_{GG}^{(0)}(n)
\right. \nonu \left.\pm\sqrt{(\gamma_{qq}^{(0)}(n)-\gamma_{GG}^{(0)}(n))^2+
4\gamma_{qG}^{(0)}(n)\gamma_{Gq}^{(0)}(n)}\right]~~, 
\ee
one can write  the $2\times 2$ matrix  in terms of  projectors and   eigenstates as follows 

\be\label{gammadec}
\Gamma^{(0)}(n)=\gamma_+(n)\mathcal{P}_+(n)+\gamma_-(n)\mathcal{P}_-(n)
\ee
with
\be\label{projdef}
\mathcal{P}_\pm(n)=\frac{\pm1}{\gamma_+(n)-\gamma_-(n)}[\Gamma^{(0)}(n)-\gamma_\mp(n)~{\rm I}]
\ee
They fulfills the usual projector properties, i.e. 
\be
\mathcal{P}_++\mathcal{P}_-=1\nonu
\mathcal{P}_\pm^2=\mathcal{P}_\pm\nonu
\mathcal{P}_+\mathcal{P}_-=\mathcal{P}_-\mathcal{P}_+=0
\ee

Solutions of  Eqs. \eqref{MOMNS} and \eqref{MOMS} are given by
\be
M_{NS}(n,\mu)=\left[\frac{\alpha_s^{LO}(\mu,N_f)}
{\alpha_s^{LO}(\mu_0,N_f)}
\right]^{[\gamma^{(0)}_{NS}(n)/2\beta_0(N_f)]}
\nonu \times ~M_{NS}(n,\mu_0)\label{alphaM}\\&&
\overrightarrow{M}(n,\mu)=\left[\frac{\alpha_s^{LO}(\mu,N_f)}
{\alpha_s^{LO}(\mu_0,N_f)
}\right]^{[\Gamma^{(0)}(n)/2\beta_0(N_f)]} \nonu \times ~
\overrightarrow{M}(n,\mu_0)\label{alphaMS}
\ee
Notably,  Eq. (\ref{alphaMS}) can be put in a more simple form by using the eigenvalues,
 $\gamma_\pm$,  and the corresponding projectors $\mathcal{P}_\pm$, 
 viz.  \cite{anodimrad} 
\be
\overrightarrow{M}(n,\mu)=
\Bigl\{\Bigl[\frac{\alpha_s^{LO}(\mu,N_f)}{\alpha_s^{LO}(\mu_0,N_f)}
\Bigr]^{[\gamma_+(n)/2\beta_0]}
\mathcal{P}_+ \nonu
 +\Bigl[\frac{\alpha_s^{LO}(\mu,N_f)}{\alpha_s^{LO}(\mu_0,N_f)}\Bigr]^{[\gamma_-(n)/2\beta_0]}
\mathcal{P}_-\Bigr\}  ~
\overrightarrow{M}(n,\mu_0).
\ee
Indeed, we are interested to actually evolve only the moment  $n=1$ of $f_{NS}$, since for this
moment we can find several lattice calculations 
(but using different approximations; cf Sec. \ref{results}). 
Our procedure requests to backward-evolve the lattice $M^{LAT}_{NS}(1,\mu_{LAT})$, down to a scale 
$\mu_0$ where 
$M^{LAT}_{NS}(1,\mu_0)=M^{CCQM}_{NS}(1)$ (notice the absence of the unknown 
scale dependence in the CCQM first moment).
 This value of the scale  is taken as $\mu_{CCQ}$.
From Eq. \eqref{alphaM}, one recognizes the necessity  to first  determine
 $\alpha^{LO}_s(\mu_{LAT},N_f)$. This can be accomplished
starting
from a reasonable value of $\alpha_s(\mu_i,3)$, like $\alpha_s(\mu_i=1~GeV,3)=0.68183$ given  
in Ref. \cite{martin} (see  also Sect. \ref{results} for the quantitative elaboration).  
 To perform this step
we have used the well-known expression
\be\label{alfamu2}
\alpha_s^{LO}(\mu,N_f)=\frac{\alpha_s^{LO}(\mu_i,N_f)}
{1+{\alpha_s^{LO}(\mu_i,N_f)\over 4 \pi}\beta_0(N_f)~\ln\Bigl(\mu^2/\mu_i^2\Bigr)}
\ee
It should be pointed out that, as explicitly shown in Eqs. \eqref{beta0} and \eqref{alfamu2}, 
$\alpha_s(\mu,N_f)$ depends upon the number of flavors $N_f$, at a given scale.
 Indeed, one has to be particularly careful about the energy scales involved, 
 when one moves from a relatively low $\mu_i=1~{\rm GeV}$ to  $\mu_{LAT}=2~{\rm GeV}$, large enough to produce a
 new quark flavor, so that  $N_f$ increases from $3$ to $4$. 
 In practice, a  two-step procedure has been adopted  for moving from 
  $\alpha_s(\mu_i,N_f=3)$ to  $\alpha_s(\mu_{LAT},N_f=4)$, by properly 
  changing $\beta_0(N_f)$, at the threshold $\mu=m_c$, i.e the mass of the
  charm.

In what follows, it is also useful to define, at a given energy scale and number of flavors,
\be
\ln (\Lambda^{N_f}_{QCD})=\ln(\mu)-{ 2\pi \over \beta_0(N_f)~\alpha_s^{LO}(\mu,N_f)}
\label{lamqcd}\ee
\subsection{QCD Evolution of GFFs}\label{Evogff}
 Similarly to the more familiar case of PDFs, where the QCD interaction among partons lead to 
 ultraviolet divergences which are
  factored out and absorbed into a dependence upon the energy scale, also in the case of GPDs one has to deal with 
  the issue of finding and solving evolution equations. As a matter of fact, GPDs do not depend on three variables but on four, namely
  $H(x,\xi, t, \mu)$ and $E(x,\xi, t, \mu)$. However,  the evolution kernel does not depend on $t$, so that the relevant variables
   for the evolution are $x$, $\xi$, and $\mu$. One should keep in mind that the evolution of GPDs is produced by 
   the combination of two regimes: (i) the one pertaining to the valence region ($|x|>|\xi|$) and 
   (ii) the one pertaining to the non valence region ($|x|<|\xi|$). One could roughly say that the evolution of
    GPDs {\em interpolates} 
   \cite{Diehlpr} between the two regions and therefore the evolution kernel has to take into account the 
   suitable physical content. 
   In particular, in the valence region a kernel  acts with a structure like the one 
   present in the DGLAP
   equations, while in the non valence region a   modified 
   ERBL kernel  is involved (see Refs. \cite{ji97}  and \cite{TGPDevol} for details on the
   evolution of vector  and tensor GPDs, respectively). 

In our actual comparison, we do not consider the full GPDs, but rather their Mellin moments, since they
  can be in principle addressed
by the lattice calculations.  As a matter of fact, GFFs covariantly parametrize 
 the Mellin moments of GPD (see Eqs. \eqref{VecGFF} and \eqref{TGFF}), and 
 evolve through a suitable generalization of the Eqs. \eqref{alphaM}
  and \eqref{alphaMS} (see Refs. \cite{anodimrad,Broni08b,Aevol,Kivel,Kirch,Broni10,Broni11}). Let us recall, however, 
  that GFFs are the coefficients 
  of  polynomials in $\xi$ that yield the Mellin moments of GPDs and not the
   Mellin moments themselves: for this reason in general the equations 
   describing GFFs evolution are more complicated than Eqs. \eqref{alphaM} and \eqref{alphaMS}.
    Indeed one can find some notable exceptions where the equations have a simple multiplicative  structure.

For the vector GFFs $A^I_{ni}(t,\mu^2)$, one should recall  
that the evolution of the isoscalar (singlet) GPD, and consequently the evolution 
of  the corresponding Mellin moments, is coupled with the evolution of the gluonic component.
This leads one to separate the evolution of GFFs 
with even and odd $n$, since for symmetry reasons  the even GFFs come
 from the isoscalar  GPDs, while the odd ones
come from the isovector  GPDs.
By repeating  the main steps given in Ref. \cite{Aevol} (see also
\cite{Kivel,Kirch} where general discussions are presented) for obtaining   the evolution equation
of both non singlet and singlet vector GFFs, we can express the results in Ref.
 \cite{Aevol}
also as follows  
\be\label{geneqA}
A_{2k+1,2\ell}(t,\mu)={\Gamma(2k+1)\over 2}~\sum_{j=k-\ell}^k~2^{2(j-k)}
\nonu
\times ~\sum_{m=j}^k~(4m+3)L_{2m+1}~{(-1)^{m-j}~\Gamma(j+m+3/2)\over \Gamma(2j+1)}
\nonu
\times ~{A_{2j+1,2(j-k+\ell)}(t,\mu_0) \over\Gamma(m-j+1)\Gamma(k-m+1)\Gamma(k+m+5/2)}
\ee
with $0\le \ell\le k$ and
\be
L_{2m+1}=\left(\frac{\alpha_s(\mu,N_f)}{\alpha_s(\mu_0,N_f)}\right)^{[\gamma^{(0)}_{qq}(2m)/2\beta_0(N_f)]}~~.
\ee
For the singlet vector GFFs  we get
\be\label{geneqA1}
A_{2k+2,2\ell}(t,\mu)=\Gamma(2k+2)~\sum_{j=k-\ell}^k ~2^{2(j-k)-1}\nonu
\times ~\sum_{m=j}^k(4m+5)L_{2m+2}~{ (-1)^{m-j}~\Gamma(j+m+5/2)\over \Gamma(2j+2)}
\nonu \times ~{A_{2(j+1),2(j-k+\ell)}(t,\mu_0)\over \Gamma(m-j+1)\Gamma(k-m+1)\Gamma(k+m+7/2)}
\ee
with $0\le \ell\le k+1$, $A_{00}=0$ and 
\be
L_{2m+2}=\left(\frac{\alpha_s(\mu,N_f)}{\alpha_s(\mu_0,N_f)}\right)^{[\Gamma^{(0)}_V(2m+1)/2\beta_0(N_f)]}\label{L2evol}
\ee
where $\Gamma^{(0)}_V$ is  the same $2\times 2$ matrix defined in \eqref{anodimatrix} 
that depends upon $N_f$ through 
$\gamma^{(0)}_{GG}$. In Eqs. \eqref{geneqA} and 
\eqref{geneqA1}, $\Gamma(k)$ is the usual Euler function. Notice that   the dependence upon 
$t$ in the GFF
  is not involved in the evolution. Some example of explicit evolution equations are
\be
A_{n0}(t,\mu)=L_{n}~A_{n0}(t,\mu_0)\nonu
A_{22}(t,\mu)=L_2~A_{22}(t,\mu_0)
\ee
where  $A_{22}(t,\mu)$ is a 2D vector, with quark and gluon components, and $L_2$ a $2\times 2$ matrix.
For  $ A_{n0}(t,\mu)$ the evolution equations become exactly equal to 
Eqs. (\ref{alphaM}) and (\ref{alphaMS}) for the odd and even $n$'s respectively. Moreover, 
 since $\gamma^{(0)}_{qq}(n=0)=0$ then $L_1=1$. This is
 expected since $A_{10}(t=0)$ is the charge ($A_{10}(t)$ is the em form factor), namely
 a measurable quantity and therefore it cannot evolve. 

For the tensor GFF $B_{ni}^I(t,\mu)$,  analogous arguments can be carried out, but
 with a 
great simplification. In fact, at LO the gluon-quark and quark-gluon transition
 amplitudes that lead to the corresponding  splitting functions are vanishing 
 for the helicity conservation (recall that $E_T (x,\xi,t)$ is related to an 
 expectation value with a transversely polarized quark and describes helicity flip 
 transitions), therefore the anomalous dimension matrix $\Gamma^{(0)}_T(n)$ is 
 diagonal. Consequently, at LO it is not necessary to separate the case of even and 
 odd $n$, since there is no mixing between quark and gluon evolutions, and 
  one can eventually write the evolution equation in a form analogous to
  Eq. \eqref{geneqA}. 
In particular,  the quark component of the 
transverse GFFs, $B^{q}_{n0}(t,\mu)$,  evolves multiplicatively (see, e.g.,
\cite{anodimrad,Broni10,Broni11,Nam11}), viz
\be\label{B0evol}
B^q_{n0}(t,\mu)=L_n^{qT}~B^q_{n0}(t,\mu_0).
\ee
where
\be
L^{qT}_n=\left(\frac{\alpha_s(\mu,N_f)}{\alpha_s(\mu_0,N_f)}
\right)^{[\gamma_{qqT}^{(0)}(n-1)/2\beta_0(N_f)]}
\ee
with a transverse anomalous dimension (notice a factor of 2 difference with  
\cite{anodimrad}, due to the different normalization) given by 
\be\label{trasanodim}
\gamma^{(0)}_{qqT}(\ell)= ~-{8\over 3}~\Bigl[3-4\sum_{k=1}^{\ell+1}\frac{1}{k}\Bigr]~~.
\ee
For the sake of completeness, the gluon transverse LO anomalous dimension reads
\be\label{trasanodimg}
\gamma^{(0)}_{ggT}(\ell)= ~-6~\Bigl[{\beta_0(N_f)\over 3}-4\sum_{k=1}^{\ell+1}\frac{1}{k}\Bigr]~~.
\ee
\section {Numerical Results}\label{results}
\begin{figure}[h!]
\begin{center}
\includegraphics[width=9cm]{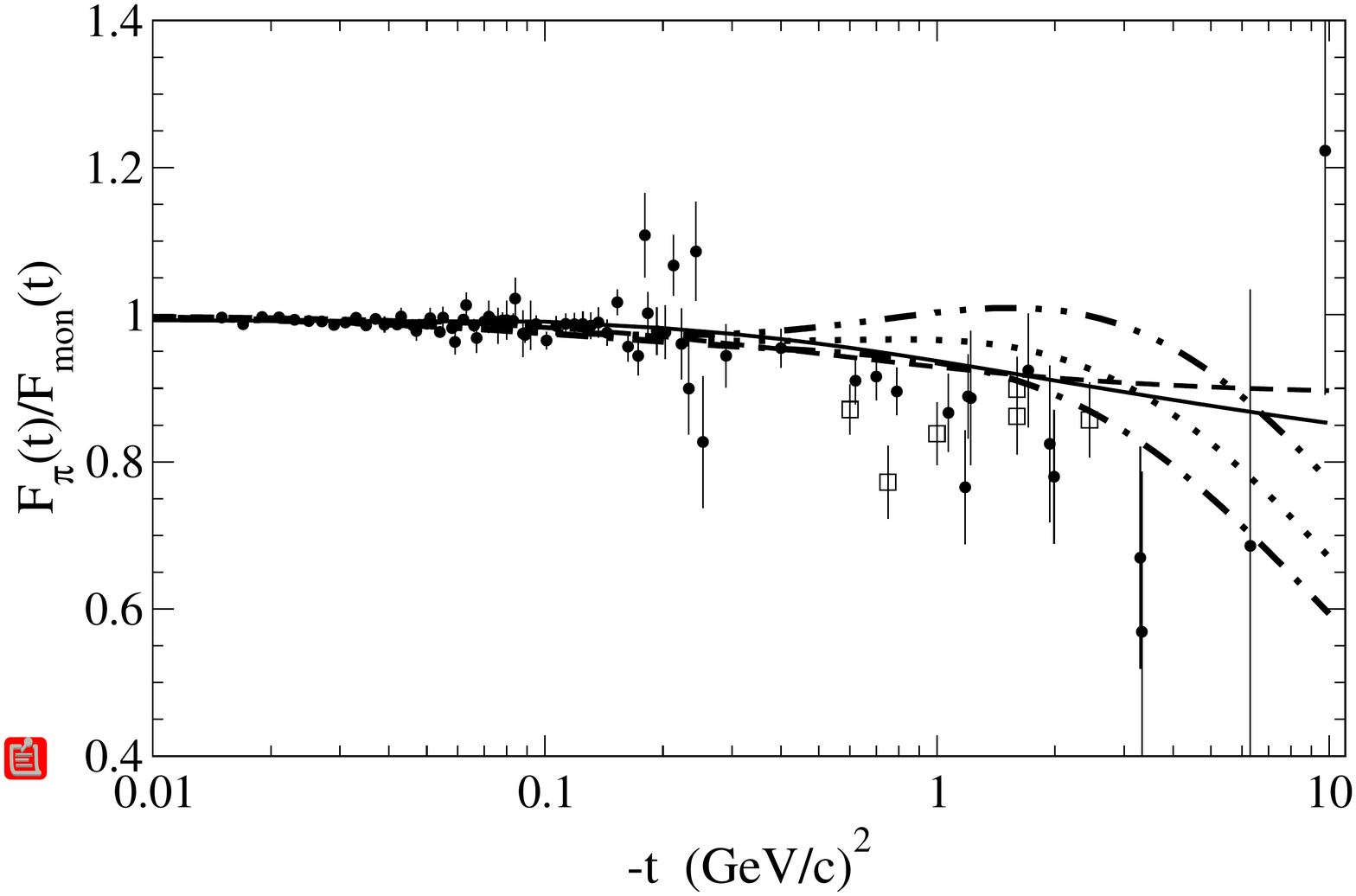}
\caption{Charged pion form factor vs $-t$. 
Solid line: LF Constituent Quark Model with $m_q=0.265$ GeV 
\cite{Melo04,Melo06}. Dashed line: monopole fit to lattice data 
extrapolated to $m_\pi=0.140$ GeV as obtained in \cite{Brom07}, 
arbitrarily extended in this figure from $-4$ to $-10\, \textrm{GeV}^2$. 
Dot-dashed line: CCQM, cf  Eq. \eqref{vertexp}, corresponding to a CQ mass 
$m_q=~0.220$ GeV and $m_R=~1.192$ GeV in the vertex
\eqref{vertexp} (recall that $m_R$ is obtained by fitting $f^{exp}_\pi=0.0922$ GeV
\cite{PDG14}).
 Dotted line: as the dot-dashed one, but with $m_q=0.210$
  GeV and $m_R=1.320$ GeV. Double-dot-dashed line: as the 
   dot-dashed one, but with $m_q=0.200$ GeV and $m_R=1.453$ GeV. Experimental data: 
  as quoted in \cite{Frede09}.}\label{Fvst}
\end{center}
\end{figure}

The reliability of the quark-pion
 vertex \eqref{vertexp}, introduced for obtaining the Bethe-Salpeter amplitude  
 \eqref{bsa}, has been first checked by comparing our
  results for a charged pion with  the most accurate 
 experimental data not affected by the evolution, i.e. 
 the spacelike em form factor. 
 Theoretically, the em form factor, $F_\pi(t)$, is given  by the GFF  
$A_{10}(t)$.
In Fig. \eqref{Fvst} the  results obtained by different models 
for the em FF are shown as a function of $(-t)$, together with the
experimental data.
 To avoid the use of a log plot, that prevents a detailed analysis, 
 the FF has been divided by the monopole function
\be
F_{mon}(t)=\frac{1}{1+|t|/m_\rho^2}
\ee
where $m_\rho=0.770$ GeV. Interestingly, in order to test the dependence 
upon the CQ mass, the results 
of our CCQM evaluated for  quark masses 
$m_q=0.200$, $0.210$, $0.220$ GeV, have been also shown, together with (i) 
the results from a LF CQM where $m_q=0.265$ GeV and a dressed 
quark-photon vertex 
were adopted \cite{Melo04,Melo06};  
(ii) a fit to the lattice data  obtained in \cite{Brom07}.
 It has to be pointed out that the fit to the lattice data was
presented in 
 \cite{Brom07} itself, and it has the following 
 monopole expression
\be
F^{lat}_\pi(t)=\frac{1}{1-t/M^2(m_\pi^{phys})}
\label{flat}\ee
with $M(m_\pi^{phys})=0.727$ GeV. The lattice  results were actually obtained for
 a pion mass  $m_\pi=0.600$ GeV, and then 
 extrapolated to the
 physical pion mass  $m_\pi^{phys}=0.140$ GeV, up to $t=-4~ \textrm{GeV}^2$ 
 (see \cite{Brom07}). In  Fig. \eqref{Fvst}, for the sake of presentation,
  the monopole fit  \eqref{flat} has been arbitrarily extended up to $t=-10\,\textrm{GeV}^2$.
 Once the CQ mass is assigned, our CCQM model depends upon only one free parameter, the regulator mass $m_R$ in Eq.
  \eqref{vertexp}.
 The value of $m_R$ is fixed by calculating the pion decay constant 
 $f_\pi$ (cf Eq. \eqref{fpi}), while the constant $C$ is determined through $F_\pi(0)$, as already
 mentioned in subsec. \ref{SecBSA}. 
 The PDG experimental  value $f^{exp}_\pi=~0.0922$ GeV \cite{PDG14}
 has been adopted. In particular, for $m_q=0.200,~210,~220$ GeV, we got 
 $m_R=~1.453$,$~1.320$,$~1.192$ GeV, respectively.

The following comments are in order: (i) a nice agreement between the CCQM
 results and the experimental FF at low momentum
transfer leads to reproduce the experimental value of the charge radius,
 $<r^2_{exp}>=0.67\pm0.02$ fm; (ii)  beyond $-t=~0.5~{\rm GeV}^{\rm 2}$,
  CCQM results begin to reveal an interesting  sensitivity  
 upon the CQ mass, 
since  if one  changes the CQ mass  by a 5\%  
then  the corresponding CCQM FF
 changes by  10-15\%, at high momentum transfer; (iii) the CCQM FFs seem to have a similar curvature of the
 data at high momentum transfer, but in order to draw reliable conclusions,
 useful for extracting information (and then improving the CCQM), it is necessary to have more
 accurate data, for  $-t\geq 1~{\rm GeV}^2$.
 
Another interesting data set  to be compared with is given by the photon-pion transition form factor, $F^*_\pi(-t)$, measured in the
process $\gamma \gamma^*\to \pi^0$
 \cite{BABAR,BELLE}. Within CCQM,
only the LO asymptotic value  of such transition FF 
can be evaluated without adding new ingredients (see, e.g., Ref. \cite{deTB11}  for a wide discussion and references quoted
therein).
 As a matter 
of fact,   one  gets for high $(-t)$, at LO in pQCD,
\be
(-t)~F^*_\pi(-t) \to {2 f_\pi\over 3}~\int_0^1 d\xi~ {\phi_\pi(\xi,|t|) \over \xi}
\ee
where $\phi_\pi(\xi,|t|)$ is the pion DA evaluated at  the scale $|t|$. The CCQM result (with an
undetermined scale for the moment, see the next subsection) is 
given by 
\be
\phi_\pi(\xi,\mu^2_{CCQM})= i{ m^2_q N_c\over f^2_\pi m_\pi(2 \pi)^2}\nonu
\times ~{1\over \xi (1-\xi)}~
 \int_0^\infty  d^2 
\kappa_\perp~ \Phi(\xi m_\pi,
\kappa_\perp)
\label{DA2}\ee
where $\Phi(\xi m_\pi,
\kappa_\perp)$ is defined in Eq. \eqref{DA1}. The normalization of $\phi_\pi$
follows from Eq. \eqref{fpi}. It should be anticipated that  CCQM results, both  
 non  evolved and evolved, as
shown in the next Fig. \ref{evolpda}, resemble 
the asymptotic pion DA obtained within the pQCD framework, i.e. $\phi^{asy}_\pi(\xi)=
6\xi (1-\xi)$, that in turn
 yields $(-t)~F^{*}_\pi(-t)\to~ 2 f_\pi$, see Refs. \cite{LB80,BL81}.
 
 In what follows, the values $m_q=0.220$ geV and $m_R=1.192$ GeV will be adopted.

\subsection{Looking for the CCQM energy scale} \label{cercamu}
As it is  well-known,  the em FF is not affected by the issue of the evolution, while 
 the other quantities we are interested in, namely the PDF and the GFFs (as well as the DA, 
 see Eq. \eqref{DA2}), have to be properly evolved. 
 
 A necessary step for going forward
  is to assign a resolution scale to CCQM. In order to perform this step, we have  taken
   lattice estimates 
  of the first Mellin moment of $f_{NS}(x,\mu)$, whose evolution is determined only by  the quark contribution,
  as normalization of our CCQM (roughly speaking). 
    The starting point is the calculation 
      of both the unpolarized GPD, 
$f_{NS}(x)=2H^{I=1}(x,0,0)$,
 and the corresponding   
Mellin moments, within  our CCQM. In particular, these quantities are shown in Table \ref{Melmo} 
up to $n=3$. 
To emphasize that  there is no direct way to gather
 information about the energy scale $\mu_0$, a  question mark is put in the Table.
\begin{table}[hbt]

\begin{center}
\caption{Mellin moments of $f_{NS}(x)$ up to $n=3$, 
 evaluated within the CCQM with the quark pion vertex given 
 in Eq. (\ref{vertexp}), $m_q=0.220$ GeV, and $m_R=1.192$ GeV. The energy 
 scale, $\mu_0$ has to be determined (see text).}\label{Melmo}
\begin{tabular}{cccc}
\hline\noalign{\smallskip}
$\mu_0$ & $<x>$ & $<x^2>$ & $<x^3>$ \\
\noalign{\smallskip}
 \hline\noalign{\smallskip}
{\bf ?} & 0.471 & 0.276 & 0.183 \\
\noalign{\smallskip}\hline
\end{tabular}
\end{center}
\end{table}
In the literature there are various lattice results for the
 first moment at the energy scale of $\mu=2$ GeV, and  
  we have exploited the ones
  shown in Tab. \ref{1mel}. It is worth noticing that the lattice results are not too far from a 
 phenomenological estimate, $<x>_{phe}(\mu=2~{\rm GeV})$, that one can deduce by  applying a 
 LO backward-evolution    to
  the value given in Ref. \cite{Holt},
     $<x>_{phe}(\mu=5.2~{\rm GeV})= 0.217(11)$, 
  obtained  after a {\em NLO re-analysis} of the Drell-Yan data of Ref. 
  \cite{Conway}. In particular, the phenomenological value at $\mu=2$ GeV is 
    $$<x>_{phe}(\mu=2~{\rm GeV})= 0.260(13)$$
\begin{table}[hbt]
\begin{center}
\caption{Recent lattice results  for the first Mellin moment of 
the non singlet $f_{NS}(x)$, at the energy scale $\mu_{LAT}= 2$ GeV. The 
first and the second lines are the results obtained   from unquenched lattice
 QCD calculations \cite{Lat07,south}, while the third 
result has been obtained in quenched lattice QCD \cite{chilf}.} \label{1mel}
\begin{tabular}{lcc}
\hline\noalign{\smallskip}
Ref. & $\mu_{LAT}[GeV]$ & $<x>_{LAT}$ \\
\noalign{\smallskip}
 \hline\noalign{\smallskip}
Lat. 07 \cite{Lat07} & 2.0 & 0.271(10) \\
South \cite{south} & 2.0 & 0.249(12) \\
$\chi$LF \cite{chilf} & 2.0 & 0.243(21) \\
\noalign{\smallskip}\hline
\end{tabular}
\end{center}
\end{table}
For the sake of completeness, it is  interesting to quote two other lattice
 calculations: (i) the quenched one of Ref. \cite{twist} that amounts to  
 a value $<x>_{LAT}(\mu=2~{\rm GeV})=0.246(15)$, i.e. falling 
between the results  of Refs. \cite{south,chilf} and (ii) 
a very recent lattice estimate, remarkably  at the
physical pion mass, giving  $<x>_{LAT}(\mu=2~{\rm GeV})=0.214(19)$ 
 \cite{twist2}.

After establishing the set of lattice data,  we need the value of $\alpha_s^{LO}$ at $\mu=2$
 GeV where $N_f=4$.
This value has been obtained starting from $\alpha_s^{LO}(\mu=1~{\rm GeV})=0.68183$ obtained in 
 Ref. \cite{martin}. Notice that at the scale $\mu=1~{\rm GeV}$
 only three flavors are active. Then, by using 
  $m_c=1.4$ GeV \cite{martin} and Eq. \eqref{alfamu2}, with the proper 
  $\beta_0(N_f)$, one    
  determines $\alpha^{LO}_s(\mu=2~{\rm GeV},4)=0.413$ 
  (see also Refs. \cite{PDG14,Marciano,MRST}, where the crossing
of the flavor threshold  has been discussed).
Finally, paying attention to the flavor threshold,    the lattice evaluations of the first moment 
  $M^{LAT}_{NS}(1,\mu_{LAT})$  have to be  backward-evolved up to a scale $\mu_0$, where
      they  match our CCQM value, 
  i.e. we  look for $\mu_0$ such that $M^{LAT}_{NS}(1,\mu_0)=0.471=M^{CCQM}_{NS}(1,?)$.  
\begin{table*}[hbt]
\begin{center}
\caption{Energy scale of CCQM, $\mu_0$, as determined from (i) the first Mellin moments calculated within a lattice
framework in Refs. 
\cite{Lat07,south,chilf} and (ii) the CCQM result, $<x>=0.471$, calculated with $m_q=0.220$ GeV and 
$m_R=1.192$ GeV. For all the three calculations shown in the Table, one gets
 $\Lambda^{N_f=3}_{QCD}(\mu_0)= 0.359$ GeV
from Eqs. \eqref{lamqcd} 
and \eqref{mu0}. }
\label{posmu}
\begin{tabular}{lccc}
\hline  \noalign{\smallskip}
Ref.  & $<x>_{LAT}$ & $\mu_0$ [GeV]& $\alpha^{LO}_s(\mu_0,3)$\\
\noalign{\smallskip}\hline\noalign{\smallskip}
Lat. 07 \cite{Lat07} & 0.271 & 0.549 & 1.64  \\
South \cite{south} & 0.249 & 0.506 & 2.04  \\
$\chi$LF \cite{chilf} & 0.243 & 0.496 & 2.17  \\
\noalign{\smallskip}\hline
\end{tabular}
\end{center}
\end{table*}
\begin{table}[hbt]
\begin{center}
\caption{Comparison for the second and third Mellin moments
 of 
the non singlet $f_{NS}(x)$, at the energy scale $\mu_{LAT}=2$ GeV, between the
unquenched lattice results of Ref. \cite{Lat07} and the evolved CCQM, where the
theoretical uncertainty is generated by the three values for the CCQM initial
scale shown in Tab. \ref{posmu}} \label{moremel}
\begin{tabular}{lcc}
\hline\noalign{\smallskip}
~ & $<x^2>$ & $<x^3>$ \\
\noalign{\smallskip}
 \hline\noalign{\smallskip}
Lat. 07 \cite{Lat07} & 0.128(18) & 0.074(27) \\
CCQM  & 0.105(11)  & 0.055(7) \\
\noalign{\smallskip}\hline
\end{tabular}
\end{center}
\end{table}
In detail, we calculate first  (cf. Eq. (\ref{alphaM})) the lattice result at the charm mass scale, viz
\be
M_{NS}(1,m_c)=\left[\frac{\alpha^{LO}_s(m_c,4)}{\alpha^{LO}_s(\mu_{LAT},4)}
\right]^{\gamma^{(0)}_{qq}(1)/(2\beta_0(4))}\nonu \times ~M_{NS}(1,\mu_{LAT})
\ee
where $\gamma^{(0)}_{qq}(1)=64/9$, $\beta_0(4)=25/3$, $\alpha_s^{LO}(m_c,4)=0.513$
(corresponding to $\Lambda_{QCD}(N_f=4)=0.322$ GeV) and $M_{NS}(1,\mu_{LAT})$ 
are the values shown in Tab. \ref{1mel}. 
Once $M_{NS}(1,m_c)$ is obtained,  $\alpha_s^{LO}(\mu_0)$ can be evaluated 
through (cf. Eq. (\ref{alphaM}))
\be
\alpha^{LO}_s(\mu_0,3)=\alpha_s^{LO}(m_c,3)\nonu \times ~\left[\frac{M_{NS}(1,\mu_0)}
{M_{NS}(1,m_c)}\right]^{-\gamma^{(0)}_{qq}(1)/(2\beta_0(3))}~,
\ee
where $M_{NS}(1,\mu_0)$ corresponds to our CCQM calculation and $\beta_0(3)=9$. After determining 
$\alpha^{LO}_s(\mu_0,3)$,  $\mu_0$ is easily found  through
\be
\ln\left({\mu_0\over \mu=1~{\rm GeV}}\right)={ 2\pi \over \beta_0(3)}\nonu
\times ~\Bigl[{1\over \alpha^{LO}_s(\mu_0,3)}-
{1\over \alpha^{LO}_s(\mu=1~{\rm GeV},3)}\Bigr]
\nonu
\label{mu0}\ee

The results for $\mu_{CCQ}$ obtained from the above   procedure, applied to the three lattice data, are shown in Tab. \ref{posmu} for $m_q=0.220$ GeV and 
$m_R= 1.192$ GeV.
In particular, the values in the third column of Tab. \ref{posmu} are used in 
the next sections as starting values for the evolution of both the non singlet 
PDF and the GFFs. The difference between the three values of $\mu_0$ in the Tab. \ref{posmu} is
 assumed as a theoretical uncertainty of our results.
To complete this subsection, in Tab. \ref{moremel},  the comparison
 with the lattice calculation of Ref. \cite{Lat07} for the second and the third
  Mellin moments is presented.

\subsection{The evolution of the non singlet PDF and the comparison with the
experimental data}
\begin{figure}[h!]
\begin{center}
\includegraphics[width=10cm]{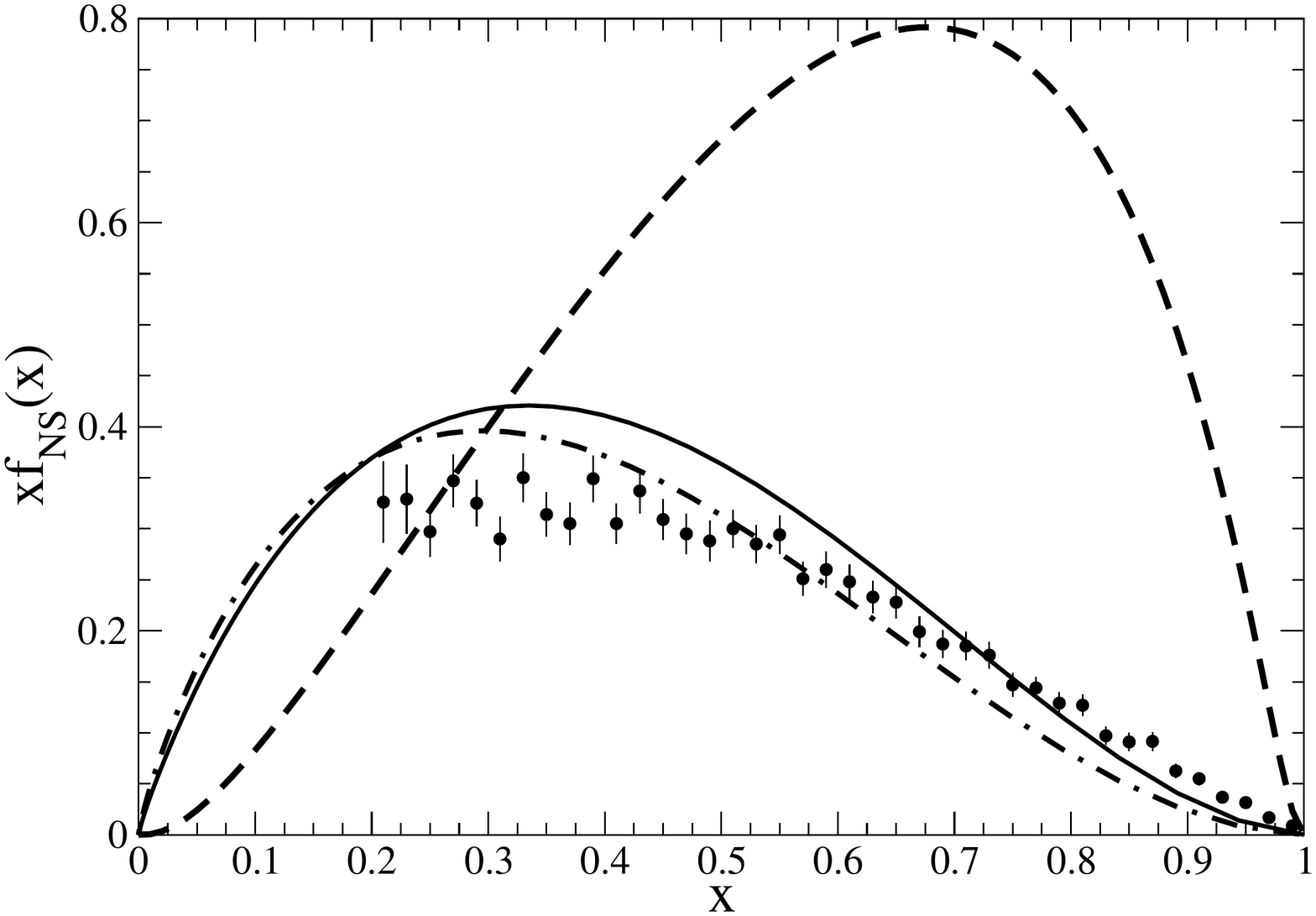}
\caption{Evolution of the non singlet parton distribution.  Dashed line: 
non evolved PDF obtained from CCQM 
$H^{I=1}(x,0,0)$ with a CQ mass $m_q=0.220$ GeV and $m_R=1.192$ GeV in the vertex
\eqref{vertexp}.  Solid line: PDF LO-evolved at $\mu=4$ GeV from $\mu_0=0.549$ GeV. 
Dot-dashed line: PDF LO-evolved at
 $\mu=4$ GeV from $\mu_0=0.496$ GeV.  For details on the 
  values of $\mu_0$ see text and Tab. \ref{posmu}. Full dots:  experimental data at the energy scale $\mu =4$ GeV,
  as given in Ref. \cite{Conway}}\label{evolpdf}
\end{center}
\end{figure}
The non singlet PDF, as already explained, is the simplest to be evolved since
 one does not need information on the gluon distribution. The  evolution has
  been performed using the FORTRAN code described in 
\cite{Kumano} that adopts a brute force method to solve the LO DGLAP equation 
for the distribution $xf_{NS}(x)$, and it  requests as input the values of (i) $\mu$, the
  final scale, 
  and (ii) the initial 
  $\Lambda^{N_f}_{QCD}$ 
   and $\mu_0$, as given  in Table \ref{posmu}.
    It should be pointed out an important detail in our calculations.
  For all the values of $\mu_0$, the evolution has been performed in two 
  steps: first   $xf^{CCQM}_{NS}(x)$ has been evolved from $\mu_0$ up to
   $m_c=1.4$ GeV and 
  then from $m_c$ up to $\mu=4$ GeV, the energy scale of the experimental 
  data \cite{Conway}. This is necessary for  taking into account the 
  variation of $N_f$, $\Lambda_{QCD}$(recall that $\Lambda^{N_f=4}_{QCD}(\mu=2~GeV)$ is  $0.322$ GeV) 
  and consequently $\alpha_s^{LO}(\mu)$.

In Fig. \ref{evolpdf}, the dashed  line is the non evolved CCQM calculation with  $m_q=0.220$ GeV and $m_R= 1.192$ GeV,
while 
the  solid and the dot-dashed lines   correspond to our evolved CCQM
 starting from  the initial scales 
$\mu_0=0.549$ GeV and  $\mu_0=0.496$ GeV, respectively. The differences between the evolved calculations
can be interpreted   as  the theoretical 
uncertainty of our calculations. However it is very interesting that our LO-evolved calculations  nicely  
 agree with the experimental data of Ref. \cite{Conway} for $x>0.5$ (see also the same
 agreement achieved within the chiral quark model of Ref. \cite{Broni08b}). 
 On the other hand, it has to be pointed out 
 that  refined calculations, like (i) the ones of  Refs. \cite{Hecht,Chang14} based on the 
 Euclidean Dyson-Schwinger equation for
 the self-energy  and (ii) the NLO calculation of Ref. \cite{Aicher} based on a soft-gluon resummation,
 underestimate the PDF tail of the experimental data from Ref. \cite{Conway}, while  agree with
   the analysis of the
 same experimental data carried out in Ref. \cite{Holt}, within a NLO framework. 
 The reanalysis of the experimental data 
 leads to
 a tail for large $x$ that has a rather  
 different derivative with respect to the original data from Ref. \cite{Conway}.
\begin{figure}[h!]
\begin{center}
\includegraphics[width=10cm]{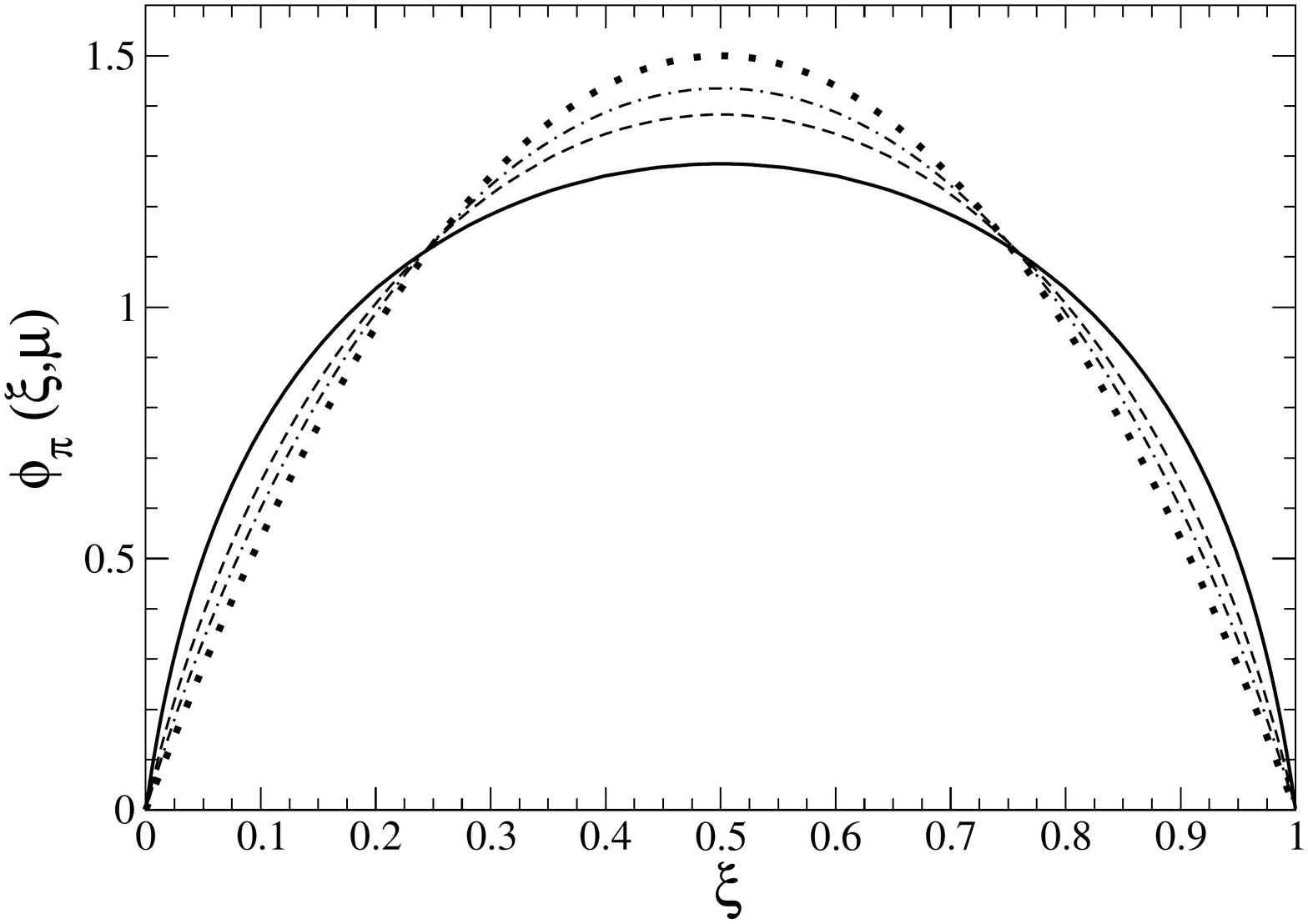}
\caption{Evolution of the pion distribution amplitude.  Solid line: 
non evolved DA obtained from our CCQM 
 with $m_q=0.220$ GeV and $m_R=1.192$ GeV in the vertex
\eqref{vertexp} (see Eqs. \eqref{DA1} and \eqref{DA2}).  Dashed line: DA LO-evolved at 
$\mu=1$ GeV. 
Dot-dashed line: PDF LO-evolved at
 $\mu=6$ GeV.  Dotted line: pQCD asymptotic DA, given by $\phi_\pi(\xi)=6\xi(1-\xi)$.
  }\label{evolpda}
\end{center}
\end{figure}
For the sake of completeness,  in Fig. \ref{evolpda}, the CCQM pion DA is presented together with the results
at the energy scale $\mu=1$ GeV and $\mu=6$ GeV. It is worth noticing that
our CCQM evolves toward 
the pQCD asymptotic pion DA $\phi_\pi(\xi)=6 ~\xi~(1-\xi)$ (see, e.g.,
\cite{LB80,BL81}) as the energy scale
increases. Analogous results are obtained within the 
 chiral quark model of Ref. \cite{Broni08}. 
 \subsection{The tensor GPD}
\label{subten}
\begin{figure*}[th]
\vspace*{-4cm}
\begin{center}
\includegraphics[width=7.5cm]{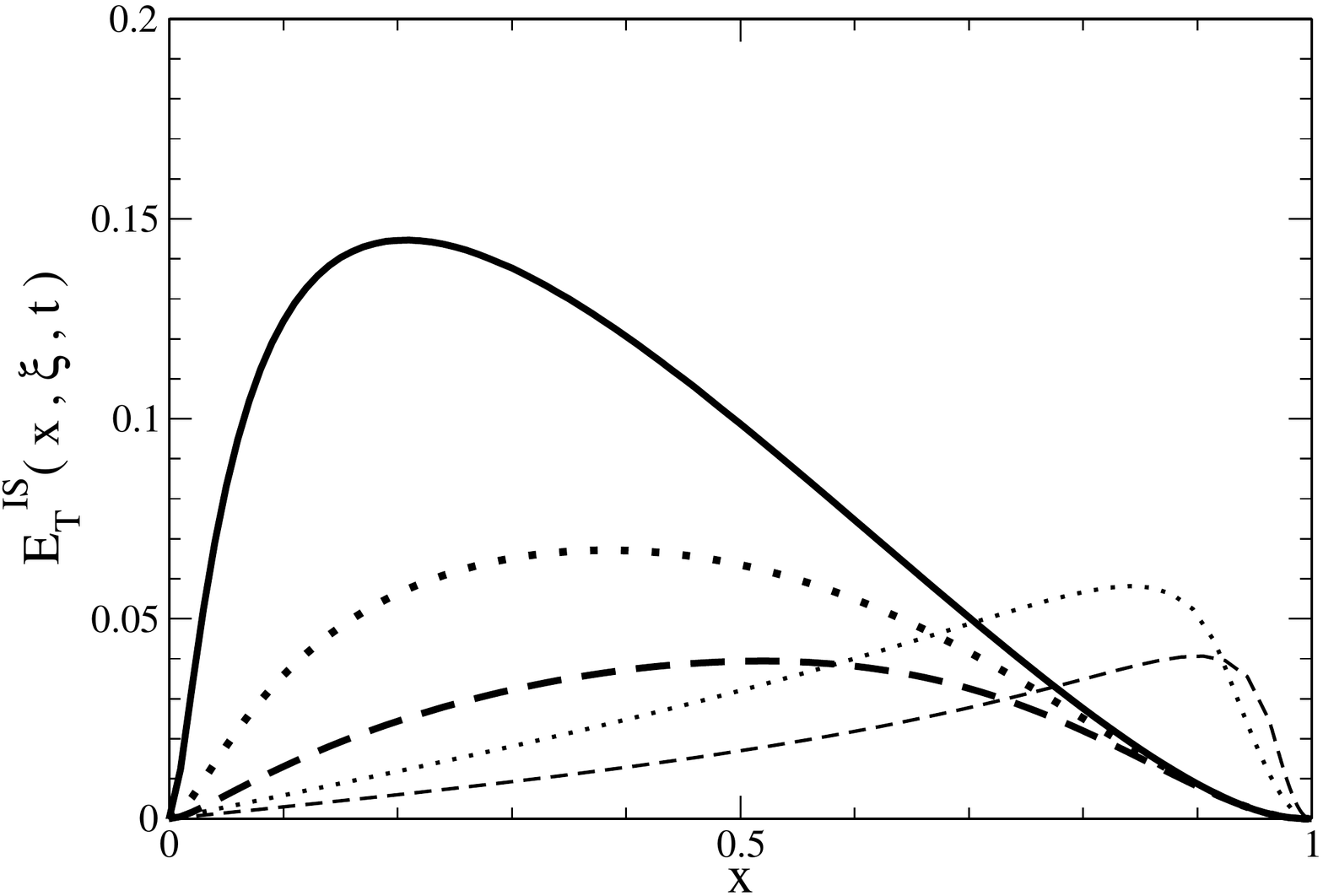} 
\hspace{0.3cm}
\includegraphics[width=7.5cm]{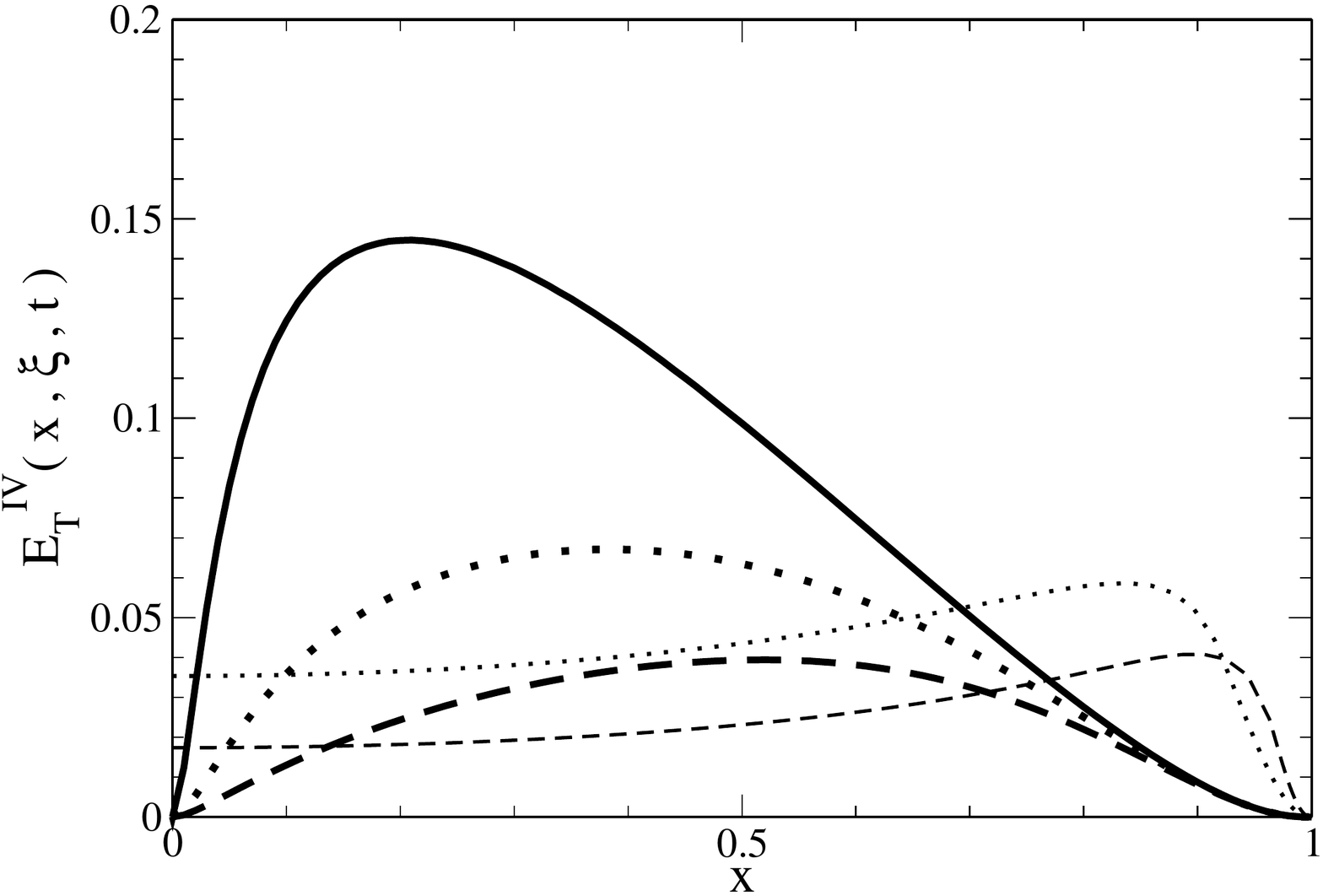}
\caption{Isoscalar and isovector tensor GPDs for a charged pion, within CCQM, for positive
$x$. The behavior for negative values of $x$ can be deduced from the antisymmetry of $E^{IS}_{\pi T}(x,\xi,t)$
and the symmetry of $E^{IV}_{\pi T}(x,\xi,t)$, respectively.
Thick solid line: $\xi=0~$ and $t=0$.
Thick dotted line: $\xi=0~$ and $t=-0.4\,GeV^2$.
Thick dashed line: $\xi=0~$ and $t=-1\,GeV^2$. Thin dotted line: 
$\xi=0.96$ and $t=-0.4\,GeV^2$. Thin dashed line:  
$\xi=0.96$ and $t=-1\,GeV^2$.}
\label{Fig_ten}\end{center}
\end{figure*}

We have extended to the tensor GPD our CCQM model already applied to the vector GPD in  
Refs. \cite{Frede09,Frede11}, and 
in Fig. \ref{Fig_ten}, our  final results 
are shown for some values of the variable $\xi$ and $t$, but for $0\leq x\leq
1$ (preliminary results were presented in Refs. \cite{Pace12,Pace13}). The GPD for negative values 
of $x$ can be obtained by exploiting the fact that $E^{IS}_{\pi T}(x,\xi,t)$ is antisymmetric 
if $x\to ~-x$, while  $E^{IV}_{\pi T}(x,\xi,t)$ is symmetric 
 (see, e.g. Ref. \cite{Diehlpr} for details). 
It has to be pointed out that for $\xi \to 0$ the valence component is dominant
(DGLAP regime) while for $\xi \to 1$ the non valence term is acting (ERBL
regime). In view of that, it is expected a peak around $x\sim 1$ for
 $\xi \to 1$, as discussed in Refs. \cite{Frede09,Frede11} for the vector GPD.  

It is worth mentioning that  both isoscalar and isovector
 tensor GPD calculated within the chiral quark model of Ref. \cite{Broni11} 
qualitatively show 
the same pattern (see also Ref. \cite{Pace12} and reference therein quoted for a
 comparison with results obtained
within the LF Hamiltonian Dynamics framework).

\subsection{The evolution of  the GFFs and the comparison with lattice data}
\label{gffevol}
The first vector GFF $A^q_{10}$, i.e. the 
em FF, is experimentally
known, while  the  other GFFs  can be investigated  only from  the theoretical 
side. In particular,  $A^q_{20}(t,\mu)$, $A^q_{22}(t,\mu)$, 
$B^q_{10}(t,\mu)$ and $B^q_{20}(t,\mu)$ have been calculated within the lattice framework
at the scale  $\mu=2$ GeV  \cite{Bromth,Brom08,Brom08b}. 
In this subsection, the comparison between our CCQM predictions and the 
above mentioned lattice evaluations is presented.
It is important to notice that other model calculations of both vector and
tensor GFFs are available in the literature (see, e.g.,
\cite{Broni10,Broni11,Nam11,Broni08,Aribron,Diehl10}).

To proceed, we have calculated  both vector and tensor GPDs, and then we have 
  extracted the relevant GFFs, by exploiting the polynomiality shown in Eqs.
\eqref{VecGFF} and \eqref{TGFF} (see also \cite{Frede09}). The main issue to 
be addressed in order to perform the
mentioned comparison with the lattice data is the evolution of our calculations up to
 $\mu_{LAT}=2$ GeV.
In the simpler case,  represented by  the tensor GFFs, 
  the LO evolution of the 
quark contribution is uncoupled from the gluon one. In particular, the two-step procedure
$\mu_{CCQ}\to m_c \to \mu_{LAT}$ has been adopted for evolving 
the two transverse GFFs, $B^q_{10}(t,\mu_0)$ and $B^q_{20}(t,\mu_0)$, 
 through Eq. \eqref{B0evol}. The needed transverse anomalous dimensions 
are given by (cf. Eq. (\ref{trasanodim}))
\be
\gamma_{qqT}^{(0)}(0)=\frac{8}{3}~~, \quad \quad \quad
\gamma_{qqT}^{(0)}(1)=8~~.
\ee
Then,   for  $\mu_{CCQ}\le \mu <  m_c$ one has $N_f=3$ and   gets
\be
B^q_{10}(t,m_c ) = B^q_{10}(t,\mu_{CCQ})
\left[\frac{\alpha^{LO}_s(m_c,3)}{\alpha_s^{LO}(\mu_{CCQ},3)}\right]^{4/27}
\label{B1evol}\\ &&
B^q_{20}(t,m_c) = B^q_{20}(t, \mu_{CCQ})
\left[\frac{\alpha^{LO}_s(m_c,3)}{\alpha_s^{LO}(\mu_{CCQ},3)}\right]^{4/9}
\label{B2evol}
\ee
 For $m_c\le \mu \le\mu_{LAT}$,  the flavor number is  $N_f=4$ and  one has
\be
B^q_{10}(t,\mu_{LAT}) = B^q_{10}(t,m_c)
\left[\frac{\alpha^{LO}_s(\mu_{LAT},4)}{\alpha_s^{LO}(m_c,4)}\right]^{4/25}
\label{B1evol1}\\ &&
B^q_{20}(t\mu_{LAT}) = B^q_{20}t,(m_c)
\left[\frac{\alpha^{LO}_s(\mu_{LAT},4)}{\alpha_s^{LO}(m_c,4)}\right]^{12/25}
\label{B2evol1}.
\ee

\begin{figure}[tbh]
\begin{center}
\includegraphics[width=9.3cm]{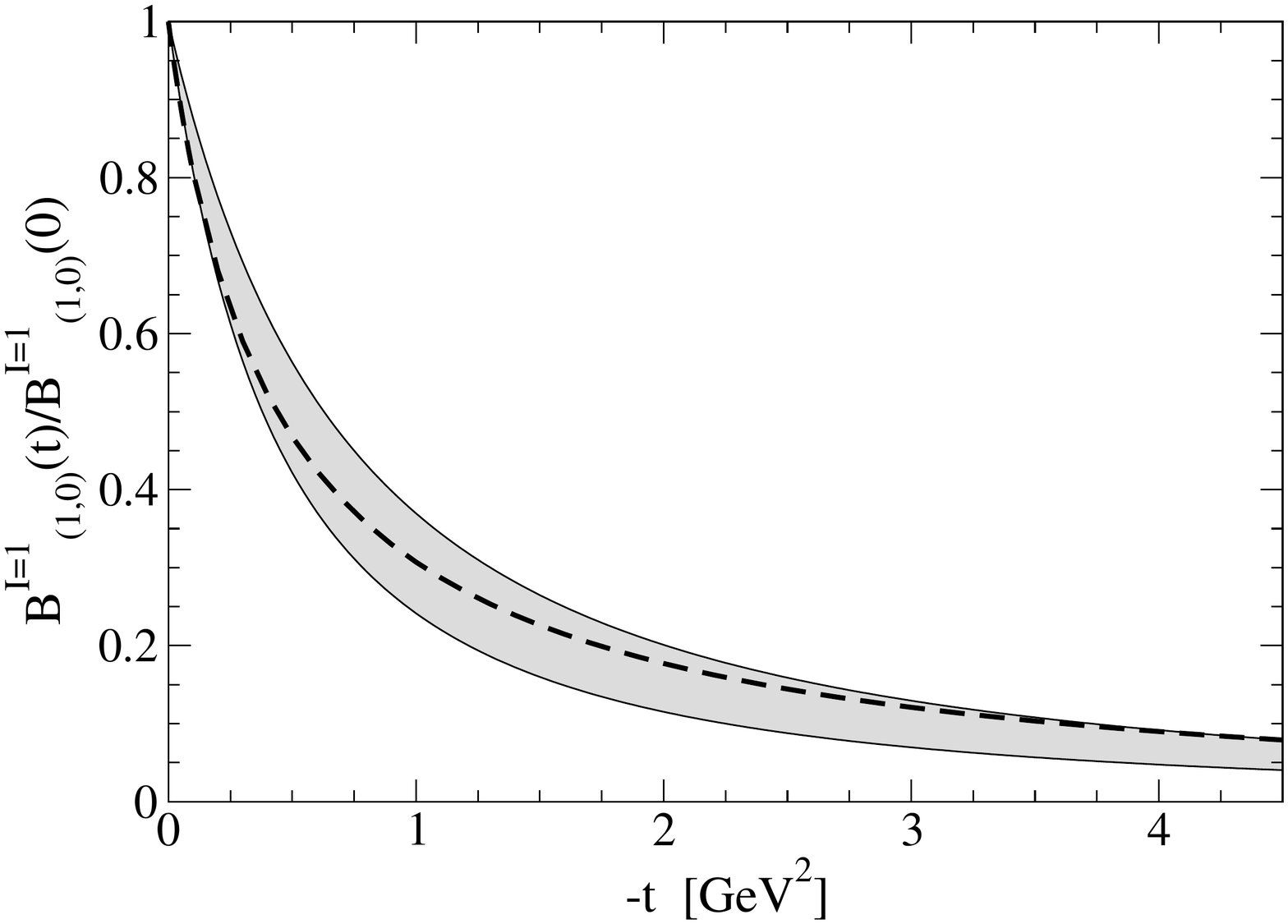}
\caption{The tensor GFF $B^q_{1,0}(t)$, an isovector one, normalized to its own charge. Dashed line: CCQM result,
corresponding to  $m_q=~0.220$ GeV and $m_R=~1.192$ GeV in the vertex
\eqref{vertexp}. The 
shaded area 
indicates the lattice data \cite{Brom08b} extrapolated to the pion physical mass $m_\pi=0.140$ GeV 
(see text).}
\label{B10g}
\end{center}
\end{figure}

\begin{figure}[tbh]
\begin{center}
\includegraphics[width=9.3cm]{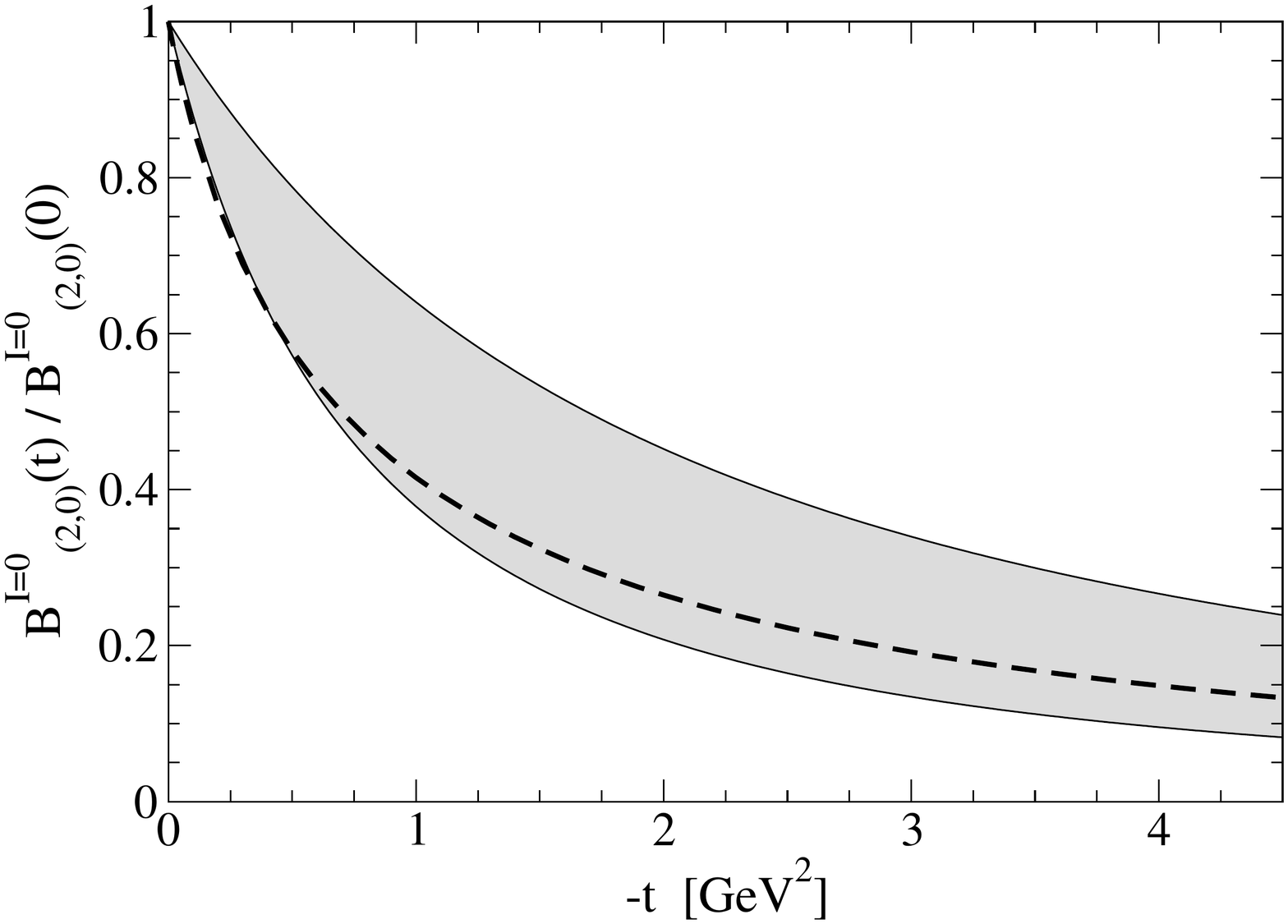}
\caption{The same as in Fig. \ref{B10g} but for the tensor GFF $B^q_{2,0}$(t).}\label{B20g}
\end{center}
\end{figure}

In the case of $A_{20}(t,\mu)$ and $A_{22}(t,\mu)$ the evolution equation 
is more complicated, since both GFFs evolve through the following expression
\be
\overrightarrow{A}_{2i}(t,\mu)=L_2\overrightarrow{A}_{2i}(t,\mu_0)
\ee
where, for both scales, one has 
\be
\overrightarrow{A}_{2i}=\left(\begin{array}{c} A^q_{2i} \\
\\
A^G_{2i}
\end{array}\right)
\ee
  From the definition (\ref{L2evol}), the exponent 
 in $L_2$ is a $2\times 2$ matrix,  (see also Eqs. \eqref{anodimatrix},   \eqref{anodim1},  \eqref{anodim2},
 \eqref{anodim3} and 
  \eqref{anodim4})
  that for $N_f=3$ reads
\be
\Gamma^{(0)}_V(1)=\left(\begin{array}{ccc}\gamma_{qq}^{(0)}(1)&~ &\gamma_{qG}^{(0)}(1)\\
&~& \\
\gamma_{Gq}^{(0)}(1)&~&\gamma_{GG}^{(0)}(1)\end{array}\right)=
\left(\begin{array}{ccc} \frac{64}{9} & ~&-\frac{2}{3}\\
 & ~ & \\
-\frac{64}{9} & ~ & 4 \end{array}\right)
\ee
with eigenvalues (see Eq. \eqref{eigen})
\be
\gamma_\pm=\frac{50\pm2\sqrt{145}}{9} ~~~.
\ee
At the valence scale,   the  gluon contribution is vanishing, and therefore 
one has $<x>_q=1/2$. Indeed, the CCQM result 
amounts to   
$<x>(\mu_{CCQ})=0.47$, namely the momentum sum rule is not completely saturated by the valence component
at the CCQM scale $\mu_{CCQ}$. This difference
 originates from the fact that we have a   covariant
description of the pion vertex, and therefore we have not only a contribution from the
 valence  LF wave function (i.e. the
amplitude of the Fock component with the lowest number of constituents), but also from components of the Fock expansion 
of the pion state
beyond the constituent one, like $|q\bar q; q \bar q\rangle$. 
Without  the gluon term at the initial scale (the assumed  valence 
one),  $A_{2i}^q(t,m_c)$ is given 
  by (cf. Eqs. \eqref{gammadec} and  \eqref{projdef})
\be\label{A2evol}
A_{2i}^q(t,m_c)=\frac{1}{2\sqrt{145}}A_{2i}^q(t,\mu_{CCQ})~{\cal R}_3^{ 25/81}
\nonu \times ~\Bigl[ (7 + \sqrt{145})~{\cal R}_3^{ \sqrt{145}/81}
-(7-\sqrt{145})~{\cal R}_3^{-\sqrt{145}/81}\Bigr]~~,
\nonu
\ee
where
\be
{\cal R}_3={\alpha_S^{LO}(m_c,3)\over\alpha_s^{LO}(\mu_{CCQ},3)}~~,
\ee
and  $A_{2i}^G(t,m_c)$ reads
\be
A_{2i}^G(t,m_c)= -~{16\over \sqrt{145}} A_{2i}^q(t,\mu_{CCQ})~{\cal R}_3^{ 25/81}
\nonu
\times ~ \Bigl[ {\cal R}_3^{\sqrt{145}/81}
-{\cal R}_3^{-\sqrt{145}/81}\Bigr]
\ee
For  $N_f=4$,  Eq. \eqref{A2evol}  changes, since both $\beta_0$ and $\gamma_{GG}^{(0)}(1)$ 
depend on the  flavor number. Therefore  $\Gamma^{(0)}_V(1)$ becomes 
\be
\Gamma^{(0)}_V(1)=\left(\begin{array}{ccc} \frac{64}{9} &~& -\frac{2}{3}\\
 & ~ & \\
-\frac{64}{9} & ~ & \frac{16}{3} \end{array}\right)
\ee
with eigenvalues 
\be
\gamma_\pm=\frac{56\pm8\sqrt{7}}{9}
\ee
Then, the evolution   in the second step from $m_c\rightarrow\mu_{LAT}=2$ GeV reads
\be\label{A2evol1}
A_{2i}^q(t,\mu)=\frac{~{\cal R}_4^{28/75}}{2\sqrt{7}}
\nonu \times\Bigl\{A_{2i}^q(t,m_c)\Bigl[(1+\sqrt{7})
{\cal R}_4^{4\sqrt{7}/75}
-(1-\sqrt{7}){\cal R}_4^{-4\sqrt{7}/75}\Bigr]
\nonu
-{3\over 4} A_{2i}^G(t,m_c)\Bigl[
{\cal R}_4^{4\sqrt{7}/75}
-{\cal R}_4^{-4\sqrt{7}/75}\Bigr]\Big\}
\ee
where
\be
{\cal R}_4={\alpha_S^{LO}(\mu,4)\over\alpha_s^{LO}(m_c,4)}~~.
\ee
It should be pointed out that the GFFs $A_{2i}^q$ evolve
multiplicatively (recall that   the evolution is not influenced by the value of $t$), given the 
absence of the gluon contribution at the valence scale, viz
\be
A_{2i}^q(t,\mu)=A_{2i}^q(t,\mu_{CCQ})~F(\mu_{CCQ},m_c,\mu)~~.
 \label{a2i}
 \ee
From Eq. \eqref{a2i}, one realizes
that the
ratio  $$A_{2i}^q(t,\mu)/A_{2i}^q(t=0,\mu)$$ ($A_{2i}^q(t=0,\mu)$ is
also called charge) can be compared with the
same ratio obtained at a different scale, e.g. at $\mu_{CCQ}$. It is understood that the same holds for the tensor GFF.
In  Figs.
\ref{B10g} and \ref{B20g},  the tensor 
GFFs $B^q_{10}(t)$ and
 $B^q_{20}(t)$, normalized to their
own charges, are shown for  both the   CCQM model, with $m_q= ~0.220$ GeV and $m_R=~1.192$ GeV, 
and the lattice framework \cite{Bromth,Brom08b}. In particular the lattice data are represented by a 
shaded area, generated 
by the envelope of
  curves that fit the lattice data with their uncertainties.  
In Refs. \cite{Bromth,Brom08b}, the lattice data have been  first extrapolated  to the physical pion mass
through a simple quadratic (in $m_\pi$) expression, and then fitted by the following  pole form
\be
\frac{GFF^{LAT}_{j}(t)}{GFF^{LAT}_{j}(0)}=\frac{1}{\Bigl[1+t/(p_j~M_j^2)\Bigr]^{p_j}}
\label{monofit}\ee
where $p_j$ and $M_j$ are pairs of adjusted parameters, shown in Tab. \ref{fitlat},
for the sake of completeness. 
 
\begin{figure}[tbh]
\begin{center}
\includegraphics[width=9.3cm]{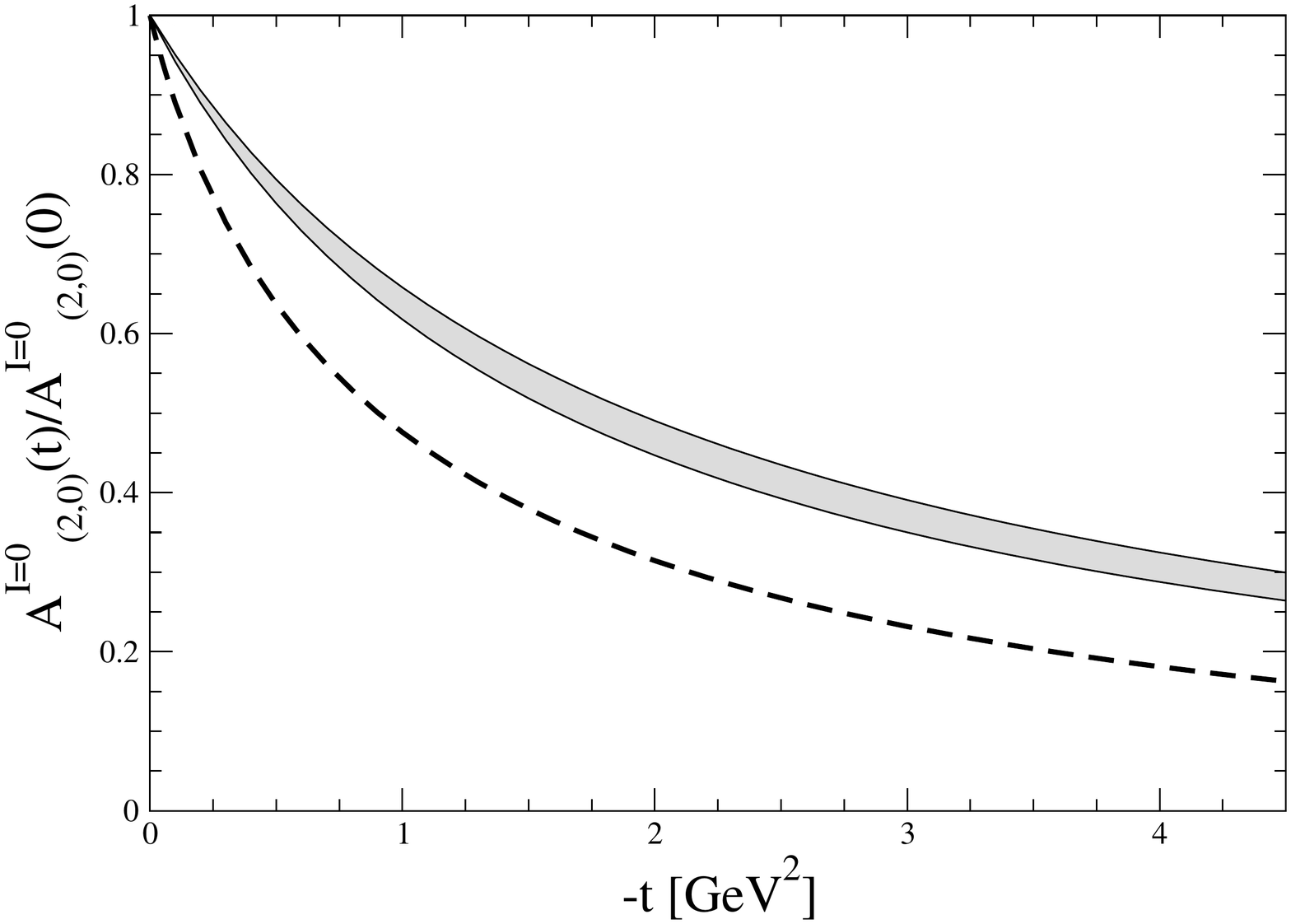}
\caption{The vector GFF $A^q_{2,0}(t)$, an isoscalar one, normalized to its own charge.  Dashed line: CCQM result,
 corresponding to  $m_q=~0.220$ GeV and $m_R=~1.192$ GeV in the vertex
\eqref{vertexp}.
The shaded area indicates the lattice data \cite{Bromth} extrapolated to the pion physical mass 
$m_\pi=0.140$ GeV (see text).}\label{A20g}
\end{center}
\end{figure}

\begin{figure}[t]
\begin{center}
\includegraphics[width=9.3cm]{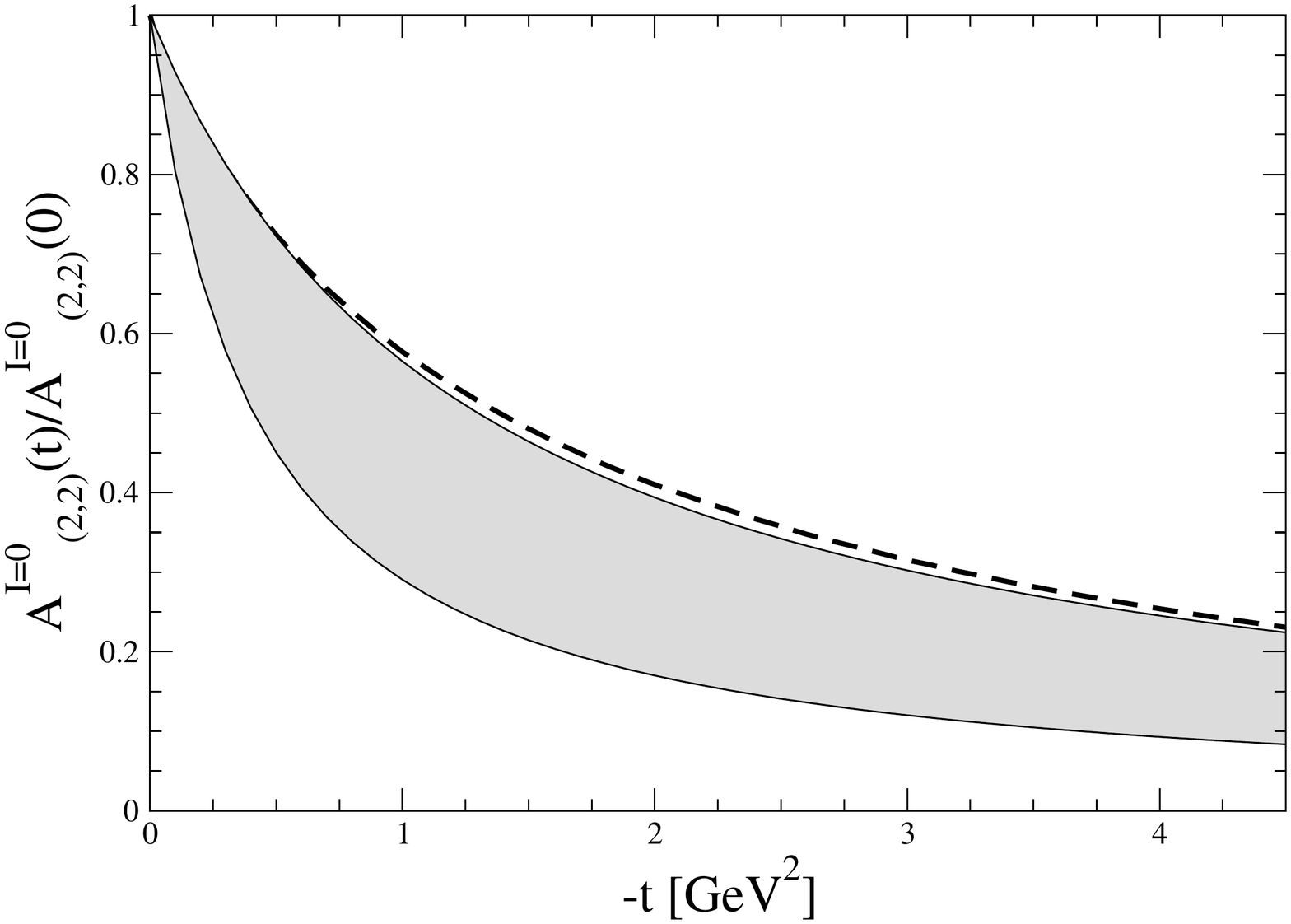}
\caption{The same as in Fig. \ref{A20g} but for the vector 
GFF $A^q_{2,2}$(t). }\label{A22g}
\end{center}
\end{figure}

In Figs. \ref{A20g} and \ref{A22g},  the  CCQM 
 $A_{2,0}(t)$ and $A_{2,2}(t)$ (with CCQM parameters different from the ones adopted 
 in Ref. \cite{Frede09}) are presented together with the corresponding lattice results.
\begin{table}[h]
\begin{center}
\caption{Adjusted parameters for describing the extrapolated lattice data
through Eq. \eqref{monofit}, as given in Refs. \cite{Bromth,Brom08b}}\label{fitlat}
\begin{tabular}{ccc}
\hline  \noalign{\smallskip}
GFF & $p_j$ & $M_j$ \\
\noalign{\smallskip}\hline\noalign{\smallskip}
$A^q_{20}(t)$ & 1 & $1.329\pm0.058$ \\
$A^q_{22}(t)$ & 1 & $0.89\pm0.25$ \\
$B^q_{10}(t)$ & 1.6 & $0.756\pm0.095$\\
$B^q_{20}(t)$ & 1.6 & $1.130\pm0.265$ \\
\noalign{\smallskip}\hline
\end{tabular}
\end{center}
\end{table}


\begin{table}[thb]
\begin{center}
\caption{Values at $t=0$ of  the CCQM GFFs, 
$A^q_{2,0}(0)$, $A^q_{2,2}(0)$, $B^q_{1,0}(0)$ and $B^q_{2,0}(0)$.}\label{GFFmu0}
\begin{tabular}{cccc}
\hline  \noalign{\smallskip}
$A^q_{2,0}(0)$ & $A^q_{2,2}(0)$ & $B^q_{1,0}(0)$ & $B^q_{2,0}(0)$ \\
\noalign{\smallskip}\hline\noalign{\smallskip}
0.4710 & -0.03308 & 0.1612 & 0.05827 \\
\noalign{\smallskip}\hline
\end{tabular}
\end{center}
\end{table}

\begin{table*}[tbh]
\begin{center}
\caption{GFFs for $t=0$ at a scale $\mu_{LAT}=2$ GeV. The first three rows contain the 
evolved (see text) CCQM results for $m_q=0.220$ GeV and $m_R=1.192$ GeV. The forth and fifth
 rows show
the  lattice extrapolations 
at the pion physical mass obtained in  Refs. \cite{Bromth,Brom08b} and in Ref. \cite{Simula},
respectively. The sixth and seventh rows present  the 
calculations from the chiral quark model 
of Refs. \cite{Broni10,Aribron} and   from
 the instanton vacuum model of Ref. \cite{Nam11}, respectively.
Notice that the results from \cite{Broni10,Aribron} were not explicitly 
written in the works, 
so that they have been extrapolated by the plots presented there.}
\label{GFFevoluti}
\begin{tabular}{lcccc}
\hline  \noalign{\smallskip}
~& $A^q_{2,0}(t=0,\mu=2 ~\textrm{GeV})$ & $A^q_{2,2}(t=0,\mu=2 ~\textrm{GeV})$ & $B^q_{1,0}(t=0,\mu=2 ~\textrm{GeV})$
  & $B^q_{2,0}(t=0,\mu=2 ~\textrm{GeV})$ \\
\noalign{\smallskip}\hline\noalign{\smallskip}
CCQM & & & &\\ \hline
$\mu_0=0.496$ & 0.2485 & -0.0175 & 0.1258 & 0.0277 \\
$\mu_0=0.506$ & 0.2542 & -0.0179 & 0.1269 & 0.0285 \\
$\mu_0=0.549$ & 0.2752 & -0.0193 & 0.1310 & 0.0313 \\
\hline
\hline
Lattice &  &  &  & \\
\hline
Ref. \cite{Bromth,Brom08b}  & $0.261\pm 0.004$ & $-0.066\pm0.008$ & $0.216\pm0.034$ & 
$0.039\pm0.010$ \\
Ref. \cite{Simula} & - & - & $0.195\pm0.010$ & - \\
\hline
\hline
 Chiral models &  &  &  &  \\ \hline
$\chi$QM \cite{Broni10,Aribron} & $0.278\pm0.015$ & - & 0.149 & 0.0287 \\
IVM \cite{Nam11} & - & - & 0.216 & 0.032 \\
\noalign{\smallskip}\hline
\end{tabular}
\end{center}
\end{table*}
If one is interested in a comparison that involves the full GFFs, then it is necessary to specify the 
scale and, accordingly, to evolve our 
 CCQM results. In particular, 
  since we have a multiplicative evolution, it is sufficient (i) to evolve only the value at $t=0$,
namely the ones
 collected in Tab. \ref{GFFmu0},  through
Eqs. (\ref{B1evol}), (\ref{B2evol}), (\ref{B1evol1}), (\ref{B2evol1}), (\ref{A2evol}) and 
 (\ref{A2evol1}) and then (ii) to use Eq. \eqref{a2i}.
  As in the case of the 
 evolution of the PDF, we considered the three possible values of $\mu_{0}$ listed in Tab. \ref{posmu}.
The results are shown in Tab. \ref{GFFevoluti}, together with lattice data \cite{Bromth,Brom08b,Simula} and model 
calculations, obtained
from a  chiral quark model \cite{Broni10,Aribron} and an instanton vacuum model \cite{Nam11}.
It should be pointed out that  within the chiral perturbation theory  (see Ref. \cite{Dono91}) one should have 
the following relation between the so-called gravitational FFs: $A^q_{22}(t)=-
A^q_{20}(t)/4 +{\cal O}(m^2_\pi)$. This relation is verified by
the lattice results, while CCQM does not. Moreover, one should  observe that 
$A^q_{20}(0)$ 
slightly
differs from $<x>$ at $\mu_{LAT}$ (see   Tab. \ref{1mel}), that contains both
quark and gluon contributions.

\begin{figure}[tbh]
\begin{center}
\hspace*{-0.2 cm}\includegraphics[width=8.7cm]{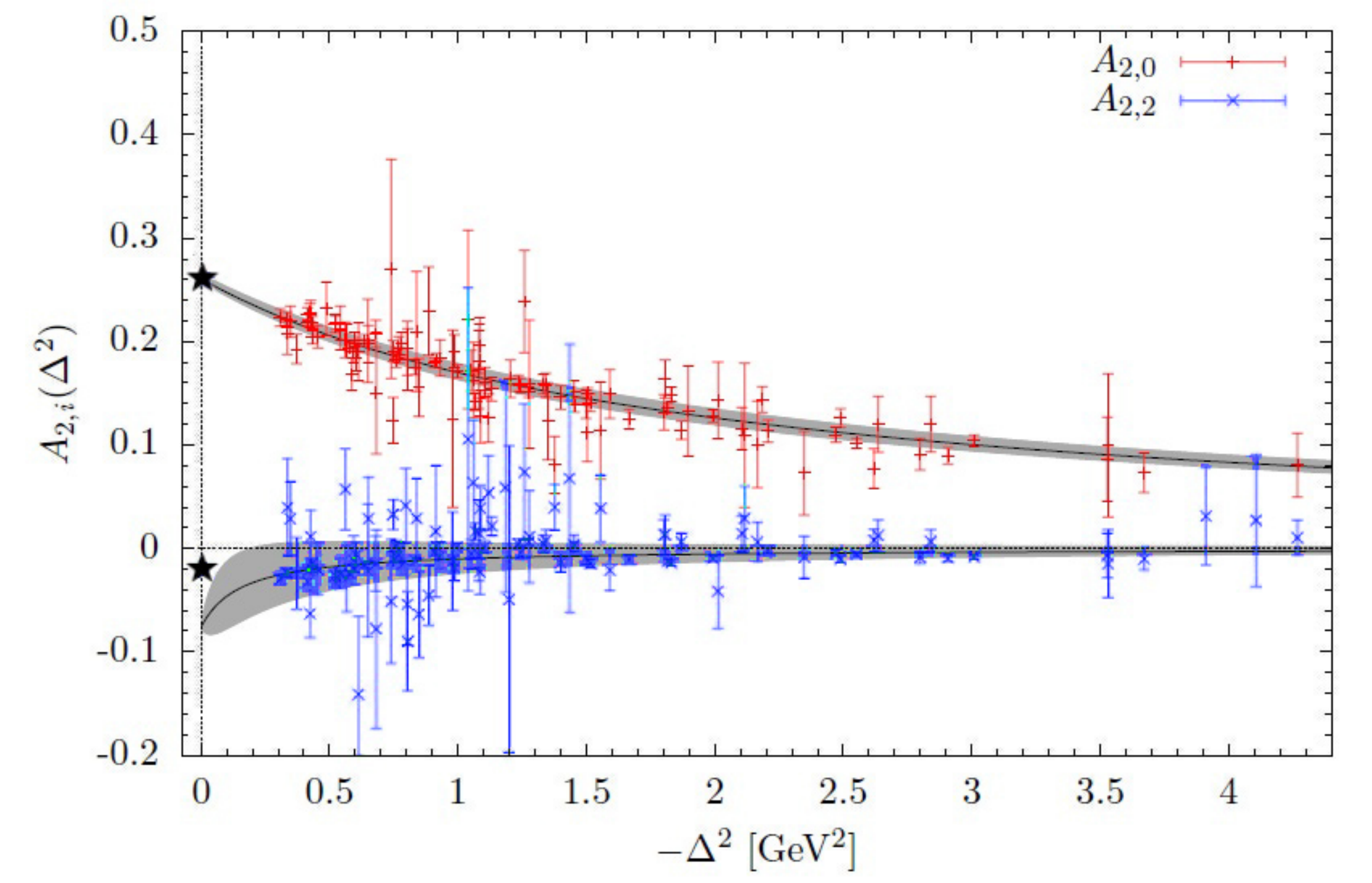}
\caption{The lattice GFFs $A^q_{2,0}(t, \mu_{LAT})$ and 
$A^q_{2,2}(t,\mu_{LAT})$ of Ref. \cite{Bromth} and the CCQM
$A^q_{2,0}(t=0, \mu_{LAT})$ and $A^q_{2,2}(t=0,\mu_{LAT})$.
 Stars:  CCQM result evolved at
$\mu_{LAT}=~2$ GeV  
(the size of the symbols is roughly proportional to the uncertainties
on CCQM initial scale $\mu_{CCQ}$).  Shaded area: uncertainties of the
 fits to the
lattice data, as estimated in Ref. \cite{Bromth} (see text). 
(Adapted from Ref. \cite{Bromth}).}
 \label{Aconfronto}
\end{center}
\end{figure}
\begin{figure}[tbh]
\begin{center}
\hspace*{-0.5 cm}\includegraphics[width=9.0cm]{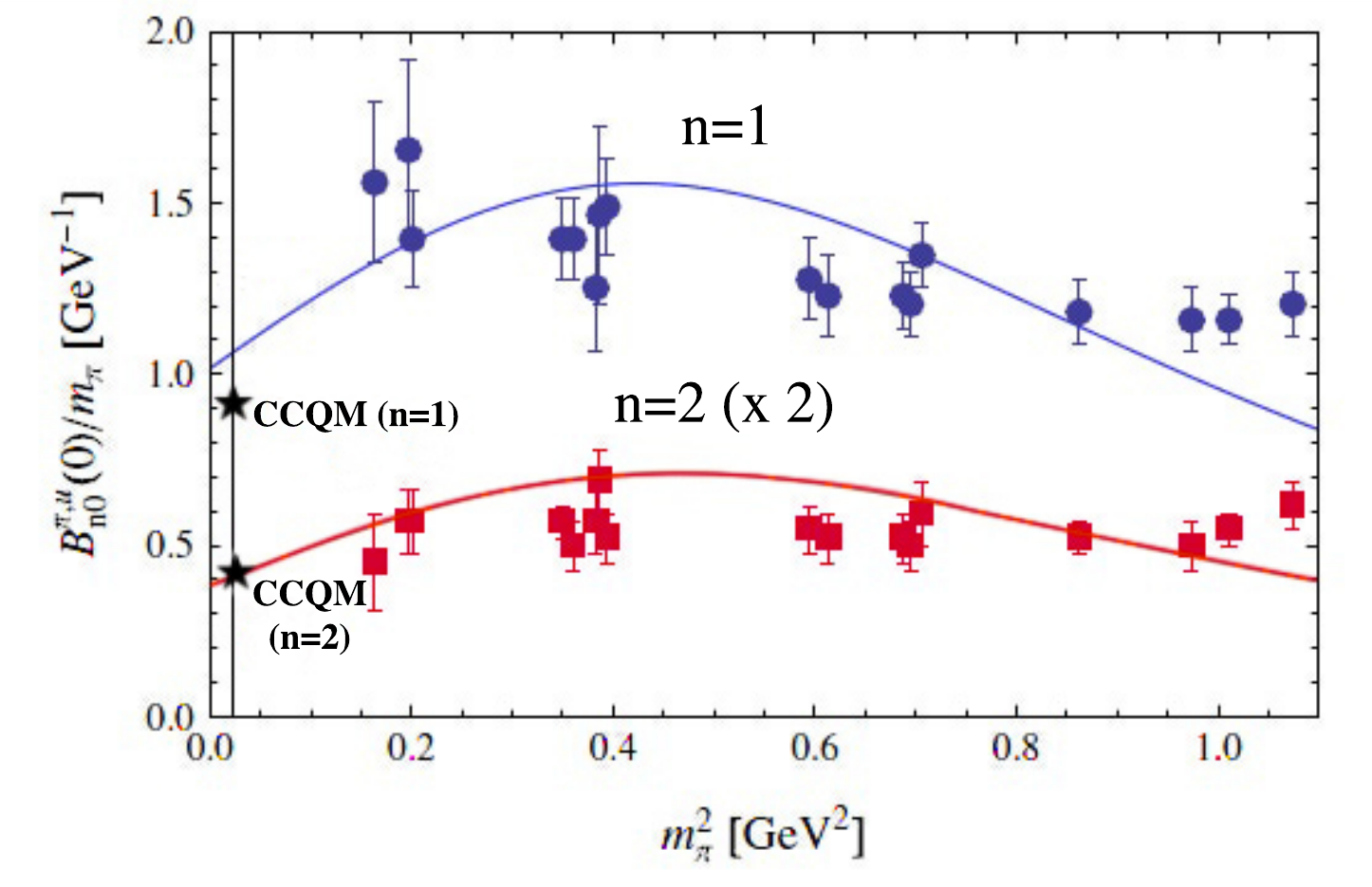}
\caption{Comparison between the CCQM $B^q_{1,0}(t=0, \mu_{LAT})$ and $B^q_{2,0}(t=0,\mu_{LAT})$, 
divided by $m_\pi$ and the
 corresponding results from the chiral quark model of Ref.
\cite{Broni10} and the lattice data of Ref. \cite{Brom08b}. Stars:  CCQM results evolved at
$\mu_{LAT}=~2$ 
 for $B^q_{1,0}(0)$ 
(upper one) and for $B^q_{2,0}(0)$ (lower one) (the size of the symbols is roughly proportional to the uncertainties
on our initial scale $\mu_{CCQ}$ as illustrated in subsec. \ref{cercamu}). Solid  lines:  results from 
the chiral quark model of Ref. \cite{Broni10}, vs $m^2_\pi$. Data points: lattice calculations from Ref. \cite{Brom08b}. 
The vertical line corresponds to the  physical pion mass.
(Adapted from Ref. \cite{Broni10}). }\label{Bconfronto}
\end{center}
\end{figure}
\begin{figure}[tbh]
\begin{center}
\includegraphics[width=10.4cm]{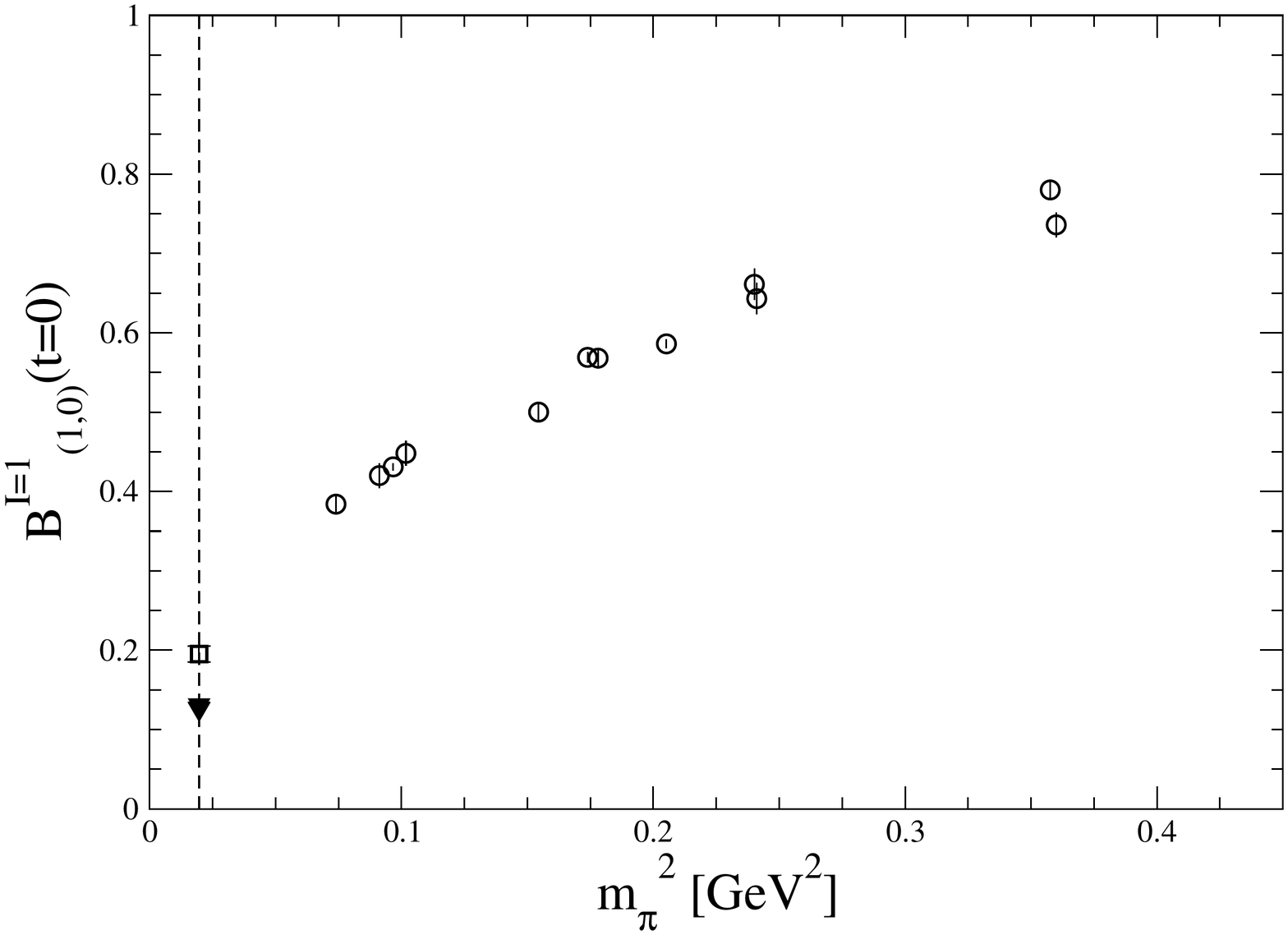}
\caption{Comparison between the CCQM $B^q_{1,0}(0)$ evolved at
$\mu_{LAT}=~2$ GeV and the lattice calculation from Ref. \cite{Simula}.  Circles: 
lattice data from \cite{Simula} for different values of
$m_\pi $.
Square: extrapolation of the previous lattice data to the  physical pion  mass, $m_\pi^{phys}$, as it has been carried out in Ref. 
\cite{Simula}. Triangle: 
CCQM result evolved at
$\mu_{LAT}=~2$ GeV (the size of the symbols is roughly proportional to the uncertainties
on our initial scale $\mu_{CCQ}$, as illustrated in subsec. \ref{cercamu}).
 (Adapted from Ref. \cite{Simula}).}
\label{B10confronto}
\end{center}
\end{figure}
To have a better understanding of the quality of the comparison  between 
our CCQM results and  the lattice data shown in Tab.
\ref{GFFevoluti}, we have added our calculation, at $t=0$,
in Fig. \ref{Aconfronto}, where 
the lattice results from \cite{Bromth}, extrapolated at the physical pion mass, 
  are presented for 
$A^q_{2,0}(t,\mu_{LAT})$ and $A^q_{2,2}(t,\mu_{LAT})$. 
In Fig. \ref{Aconfronto}, the stars  at $t=0$ represent the
CCQM  values   evolved at the lattice scale (the size of the symbols 
is roughly proportional to the uncertainties of the initial
$\mu_{CCQ}$ (cf subsec. \ref{cercamu}), while the shaded area is the
uncertainties produced by the fits to the lattice data, as elaborated in Ref.
\cite{Bromth}. It is clear that in order to have a conclusive comparison a more
wide lattice data set is necessary, but on the other hand it is impressive that
a small quantity, like $A^q_{2,2}(t,\mu_{LAT})$, can be extracted with a quite
reasonable extent of reliability. In Figs. \ref{Bconfronto} 
and \ref{B10confronto},   analogous comparisons for  $B^q_{1,0}(t=0,\mu_{LAT})$ and 
$B^q_{2,0}(t=0,\mu_{LAT})$ are shown. In particular, Fig. \ref{Bconfronto} contains
both $B_{1,0}(t=0,\mu_{LAT})$ and 
$B^q_{2,0}(t=0,\mu_{LAT})$, evaluated within CCQM (stars) and within the chiral
quark model of Ref. \cite{Broni10} with different $m_\pi$. In the figure  the
lattice data of Ref. \cite{Brom08b} are also present. Again, the comparison
between 
 values at $t=0$  and  physical pion mass
appears non trivial. In Fig. \ref{B10confronto}, a recent lattice calculation of $B^q_{1,0}(t=0,\mu_{LAT})$ \cite{Simula} is
compared with our CCQM (triangles).
In general,  one has an overall agreement,
a little bit better for $B^q_{2,0}(0)$.

The  knowledge of  GFFs allows one to investigate  the probability 
density $\rho_n({\bf b}_\perp,{\bf s}_\perp)$ for a transversely-polarized
u-quark
 (cf Eq. \eqref{rhodef}). In particular, one can address the 3D structure of 
 the pion in the impact parameter space. For instance,
 one can calculate the
average transverse shifts when the   quark is polarized along the $x$-axis, 
i.e. ${\bf s}_\perp\equiv \{1,0\}$. The  shift for a given $n$ is given by
\cite{Brom08b} \be
 \langle b_y\rangle_n= {\int d{\bf b}_\perp~b_y~ \rho_n({\bf b}_\perp,{\bf s}_\perp)
 \over \int d{\bf b}_\perp~ \rho_n({\bf b}_\perp,{\bf s}_\perp)}= {1 \over 2 m_\pi}~
 {B^q_{n,0}(t=0)\over 
 A^q_{n,0}(t=0)} \nonu
 \label{shift}
 \ee
From the CCQM values evolved at $\mu_{LAT}$, shown in Tab. \ref{GFFevoluti}, 
one 
can construct the shifts for $n=1,2$,
and then compare with the corresponding
   lattice results, as given in  
  Ref. \cite{Brom08b}. In Tab.
   \ref{tabpol}, the comparison is  shown
(recall that $A^q_{1,0}(t=0)=1$). Obviously, the same observations relevant
 for  Tab. \ref{GFFevoluti} can be  also repeated for Tab. \ref{tabpol}, since
 it contains the same information but presented in a different context. 
  \begin{table*}
  \begin{center}
\caption{Mean shifts along the direction perpendicular to the u-quark 
transverse  polarization, ${\bf s}_\perp \equiv \{1,0\}$, for $n=1,2$ (cf Eq.
\eqref{shift}). 
The CCQM results have been constructed from the values in Tab. \ref{GFFevoluti}
(notice that 
the uncertainties are originated by the three values listed there).}
\label{tabpol}
\begin{tabular}{c c c c}
\hline\noalign{\smallskip}
 ~        & CCQM - $m_\pi = 140$ MeV & lattice \cite{Brom08b}&lattice
 \cite{Simula}\\
 \noalign{\smallskip}
 \hline\noalign{\smallskip}
$\langle b_y\rangle_1$  & 0.0901 $\pm$ 0.0015 ~fm& 0.151 $\pm $ 0.024  ~fm
& 0.137 $\pm$ 0.007 ~fm\\
$\langle b_y\rangle_2$  & 0.0796  $\pm$ 0.001 ~fm  & 0.106 $\pm$ 0.028  ~fm& ~\\
\noalign{\smallskip}\hline
\end{tabular}
\end{center}
\end{table*} 
 The values shown in Tab. \ref{tabpol} indicate that even the simple version of
 a CCQM is able to reproduce a  distortion 
  of 
the transverse density
in a direction perpendicular to the quark polarization, and in turn demonstrate
the presence of  a non
trivial correlation between the orbital angular momenta  and the  spin
of the constituents inside a pseudoscalar hadron, that attracts a great interest
from both experimental and theoretical side (see, e.g., \cite{Bur05}).

\section{Conclusions}
\label{concl}

 A simple,  but fully  covariant constituent quark model  
 has been exploited  for investigating the phenomenology of the leading-order 
 Generalized Parton
 Distributions of the pion. 
 The main ingredients of the approach are (i) the
 generalization of the Mandelstam formula, applied in the seminal work of 
 Ref.  \cite{Mandel} to matrix elements
 of the em current operator between states of  a relativistic composite system, and 
 (ii) an Ansatz of the  Bethe-Salpeter amplitude  for describing the 
 quark-pion vertex. Their combination produces a very effective tool that allows
 a  careful phenomenological  investigation of the  pion, as shown in detail
 through the evaluation of both  vector and tensor pion GPDs. We have also taken
 into account, at the leading order, the evolution for obtaining a meaningful
 comparison with both experimental data (see Fig. \ref{evolpdf} for the
 comparison with the PDF extracted from the Drell-Yan data in Ref.
 \cite{Conway}) and lattice calculations of generalized form factors. 
 
Summarizing, the CCQM proves to be quite satisfactory in describing the 
pion
 phenomenology, especially considering that the model involves relatively 
 simple calculations and actually admits only one really free parameter 
 (the mass $m_q$ of the constituent quark, since $m_R$ is constrained by 
 $f_{\pi}$). It is worth noting that the CCQM is elaborated in Minkowski 
 space, and the overall agreement we have shown with the lattice data, 
 obtained in Euclidean space, could be an interesting source of information 
 on the interplay of  calculations performed in the two spaces, with a 
 particular attention to the issue of the analytic behavior. In the future 
 the present model could be substantially improved by enriching the analytic
  structure of the pion BS amplitude through a  dynamical approach based 
  on the solution of the BSE via the Nakanishi integral representation
  \cite{Naka} (cf subsect \ref{SecBSA}),
  supplemented with a phenomenological kernel. 
   In perspective, given the simplicity and the effectiveness of the approach,
   one could aim  at applying the same model to  more complex hadrons than the 
   pion, e.g. 
   to the nucleon within a quark-diquark framework.

\end{document}